\documentclass[oneside,english]{amsart}
\usepackage[T1]{fontenc}
\usepackage[latin9]{inputenc}
\usepackage{amsthm}
\usepackage{amssymb}
\usepackage{graphicx}
\usepackage{esint}

\makeatletter
\numberwithin{equation}{section}
\numberwithin{figure}{section}
  \theoremstyle{plain}
  \newtheorem*{thm*}{\protect\theoremname}
\theoremstyle{plain}
\newtheorem{thm}{\protect\theoremname}
  \theoremstyle{plain}
  \newtheorem{cor}[thm]{\protect\corollaryname}
  \theoremstyle{plain}
  \newtheorem{prop}[thm]{\protect\propositionname}
  \theoremstyle{plain}
  \newtheorem{lem}[thm]{\protect\lemmaname}
  \theoremstyle{definition}
  \newtheorem{defn}[thm]{\protect\definitionname}
  \theoremstyle{remark}
  \newtheorem{rem}[thm]{\protect\remarkname}

\makeatother

\usepackage{babel}
  \providecommand{\corollaryname}{Corollary}
  \providecommand{\definitionname}{Definition}
  \providecommand{\lemmaname}{Lemma}
  \providecommand{\propositionname}{Proposition}
  \providecommand{\remarkname}{Remark}
  \providecommand{\theoremname}{Theorem}
\providecommand{\theoremname}{Theorem}

\begin{document}

\title{The energy density in the planar Ising model}

\date{June 22nd, 2010}

\author{Clément Hongler And Stanislav Smirnov}

\address{Department of Mathematics, Columbia University. 2990 Broadway, New
York, NY 10027, USA. }

\email{\textsf{hongler@math.columbia.edu}}

\address{Section de Mathématiques, Université de Genève. 2--4 rue du Lièvre,
Case postale 64, 1211 Genève 4, Suisse. }

\email{\textsf{stanislav.smirnov@unige.ch}}
\begin{abstract}
We study the critical Ising model on the square lattice in bounded
simply connected domains with $+$ and free boundary conditions. We
relate the energy density of the model to a discrete fermionic spinor
and compute its scaling limit by discrete complex analysis methods.
As a consequence, we obtain a simple exact formula for the scaling
limit of the energy field one-point function in terms of the hyperbolic
metric. This confirms the predictions originating in physics, but
also provides a higher precision.
\end{abstract}


\maketitle
\tableofcontents{}

\section{Introduction}

\subsection{The model}

The Lenz-Ising model in two dimensions is probably one of the most
studied models for an order-disorder phase transition, exhibiting
a very rich and interesting behavior, yet well understood both from
the mathematical and physical viewpoints \cite{baxter,mccoy-wu,palmer}.

After Kramers and Wannier \cite{kramers-wannier} derived the value
of the critical temperature and Onsager \cite{onsager} analyzed the
behavior of the partition function for the Ising model on the two-dimensional
square lattice, a number of exact derivations were obtained by a variety
of methods. Thus it is often said that the 2D Ising model is \emph{exactly
solvable} or \emph{integrable}. Moreover, it has a conformally invariant
scaling limit at criticality, which allows to use \emph{Conformal
Field Theory} \emph{(CFT)} or \emph{Schramm's SLE} techniques. CFT
provides predictions for quantities like the correlation functions
of the spin or the energy fields, which in principle can then be related
to SLE.

In this paper, we obtain a rigorous, exact derivation of the one-point
function of the energy density, matching the CFT predictions \cite{di-francesco-mathieu-senechal,cardy-i,burkhard-guim}.
We exploit the integrable structure of the 2D Ising model, but in
a different way from the one employed in the classical literature.
Our approach is rather similar to Kenyon's approach to the dimer model
\cite{kenyon-i}.

We write the energy density in terms of discrete fermionic spinors
introduced in \cite{smirnov-i}. These spinors solve a discrete version
of a Riemann boundary value problem, which identifies them uniquely.
In principle, this could be used to give an exact, albeit very complicated,
formula, that one could try to simplify -- a strategy similar to most
of the earlier approaches. Instead we pass to the scaling limit, showing
that the solution to the discrete boundary value problem approximates
well its continuous counterpart, which can be easily written using
conformal maps. Thus we obtain a short expression, approximating the
energy density to the first order. Moreover, our method works in any
simply connected planar domain, and the answer is, as expected, conformally
covariant.

Similar spinors appeared in Kadanoff and Ceva \cite{kadanoff-ceva}
and in Mercat \cite{mercat}, but their scaling limits with boundary
conditions were not discussed before \cite{smirnov-i}.

Recall that the Ising model on a graph $\mathcal{G}$ is defined by
a Gibbs probability measure on configurations of $\pm1$ (or \emph{up/down})
spins located at the vertices: it is a random assignment $\left(\sigma_{x}\right)_{x\in\mathcal{V}}$
of $\pm1$ spins to the vertices $\mathcal{V}$ of $\mathcal{G}$
and the probability of a state is proportional to its Boltzmann weight
$e^{-\beta H}$, where $\beta>0$ is the inverse temperature of the
model and $H$ is the Hamiltonian, or energy, of the state $\sigma$.
In the Ising model with no external magnetic field, we have $H:=-\sum_{i\sim j}\sigma_{i}\sigma_{j}$,
where the sum is over all the pairs of adjacent vertices of $\mathcal{G}$.

\subsection{The energy density}

Let $\Omega$ be a Jordan domain and let $\Omega_{\delta}$ be a discretization
of it by a subgraph of the square grid of mesh size $\delta>0$. We
consider the Ising model on the graph $\Omega_{\delta}$ at the critical
inverse temperature $\beta_{c}=\frac{1}{2}\ln\left(\sqrt{2}+1\right)$;
on the boundary of $\Omega_{\delta}$, we may impose the value $+1$
to the spins or let them free (we call these $+$ and \emph{free}
boundary conditions respectively). Our main result about the energy
density is the following:
\begin{thm*}
Let $a\in\Omega$ and for each $\delta>0$, let $\left\langle x_{\delta},y_{\delta}\right\rangle $
be the closest edge to $a$ in $\Omega_{\delta}$. Then, as $\delta\to0$,
we have 
\begin{eqnarray*}
\mathbb{E}_{+}\left[\sigma_{x_{\delta}}\sigma_{y_{\delta}}-\frac{\sqrt{2}}{2}\right] & = & \phantom{-}\frac{l_{\Omega}\left(a\right)}{2\pi}\cdot\delta+o\left(\delta\right),\\
\mathbb{E}_{\mathrm{free}}\left[\sigma_{x_{\delta}}\sigma_{y_{\delta}}-\frac{\sqrt{2}}{2}\right] & = & -\frac{l_{\Omega}\left(a\right)}{2\pi}\cdot\delta+o\left(\delta\right),
\end{eqnarray*}
where the subscripts $+$ and \emph{free} denote the boundary conditions
and $l_{\Omega}$ is the element of the hyperbolic metric of $\Omega$.
\end{thm*}
A precise version of this theorem in terms of the energy density field
is given in Section \ref{sub:the-energy-density}. This result has
been predicted for a long time by CFT methods (see \cite{di-francesco-mathieu-senechal,burkhard-guim}
for instance), notably using Cardy's celebrated mirror image technique
\cite{cardy-i}. However, CFT does not allow to determine the lattice-specific
constant $\frac{1}{2\pi}$ appearing in front of the hyperbolic metric
element.

This is one of the first results where full conformal invariance (i.e.
not only Möbius invariance) of a correlation function for the Ising
model is actually shown. The proof does not appeal to the SLE machinery,
although the fermionic spinor that we use is very similar to the one
employed to prove convergence of Ising interfaces to SLE$\left(3\right)$
\cite{chelkak-smirnov-ii}. A generalization of our result with mixed
boundary conditions could also be used to deduce convergence to SLE.

In the case of the full plane, the energy density correlations have
been studied by Boutillier and De Tilière, using dimer model techniques
\cite{boutillier-de-tiliere-ii,boutillier-de-tiliere-i}. However,
their approach works in the infinite-volume limit or in periodic domains
and does not directly apply to arbitrary bounded domains.

In the case of the half-plane, the energy density one-point function
has been recently obtained by Assis and McCoy \cite{assis-mccoy},
using transfer matrix techniques. 

The strategy for the proof of our theorem relies mainly on:
\begin{itemize}
\item The introduction of a discrete fermionic spinor, which is a complex
deformation of a certain partition function, and of an infinite-volume
version of this spinor.
\item The expression of the energy density in terms of discrete fermionic
spinors.
\item The proof of the convergence of the discrete spinors to continuous
fermionic spinors, which are holomorphic functions.
\end{itemize}

\subsection{Graph notation\label{sub:graph-notation}}

Let us first give some general graph notation. Let $\mathcal{G}$
be a graph embedded in the complex plane $\mathbb{C}$.
\begin{itemize}
\item We denote by $\mathcal{V}_{\mathcal{G}}$ the set of the vertices
of $\mathcal{G}$, by $\mathcal{E}_{\mathcal{G}}$ the set of its
(unoriented) edges, by $\vec{\mathcal{E}}_{\mathcal{G}}$ the set
of its oriented edges.
\item We identify the vertices $\mathcal{V}_{\mathcal{G}}$ with the corresponding
points in the complex plane (since $\mathcal{G}$ is embedded). An
oriented edge is identified with the difference of the final vertex
minus the initial one.
\item Two vertices $v_{1},v_{2}\in\mathcal{V}_{\mathcal{G}}$ are said to
be \emph{adjacent} if they are the endpoints of an edge, denoted $\left\langle v_{1},v_{2}\right\rangle $,
and two distinct edges $e_{1},e_{2}\in\mathcal{E}_{\mathcal{G}}$
are said to be \emph{incident} if they share an endvertex.
\end{itemize}

\subsubsection{Discrete domains\label{sub:discrete-domains}}
\begin{itemize}
\item We denote by $\mathbb{C}_{\delta}$ the square grid of mesh size $\delta>0$.
Its vertices and edges are defined by 
\begin{eqnarray*}
\mathcal{V}_{\mathbb{C}_{\delta}} & \,:=\, & \left\{ \delta\left(j+ik\right):j,k\in\mathbb{Z}\right\} ,\\
\mathcal{E}_{\mathbb{C}_{\delta}} & \,:=\, & \left\{ \left\langle v_{1},v_{2}\right\rangle :v_{1},v_{2}\in\mathcal{V}_{\mathbb{C}_{\delta}},\left|v_{1}-v_{2}\right|=\delta\right\} .
\end{eqnarray*}

\item In order to keep the notation as simple as possible, we will only
look at finite \emph{induced} subgraphs $\Omega_{\delta}$ of $\mathbb{C}_{\delta}$
(two vertices of $\Omega_{\delta}$ are linked by an edge in $\Omega_{\delta}$
whenever they are linked in $\mathbb{C}_{\delta}$), that we will
also call \emph{discrete domains}.
\item For a discrete domain $\Omega_{\delta}$, we denote by $\Omega_{\delta}^{*}$
the \emph{dual graph} of $\Omega_{\delta}$: its vertices $\mathcal{V}_{\Omega_{\delta}^{*}}$
are the centers of the bounded faces of $\Omega_{\delta}$ and two
vertices of $\mathcal{V}_{\Omega_{\delta}^{*}}$ are linked by an
edge of $\mathcal{E}_{\Omega_{\delta}^{*}}$ if the corresponding
faces of $\Omega_{\delta}$ share an edge.
\item We denote by $\partial\mathcal{V}_{\Omega_{\delta}}$ the set of vertices
of $\mathcal{V}_{\mathbb{C}_{\delta}}\setminus\mathcal{V}_{\Omega_{\delta}}$
that are at distance $\delta$ from a vertex of $\mathcal{V}_{\Omega_{\delta}}$
(i.e. that are adjacent in $\mathbb{C}_{\delta}$ to a vertex of $\mathcal{V}_{\Omega_{\delta}}$)
and by $\partial\mathcal{E}_{\Omega_{\delta}}\subset\partial\mathcal{E}_{\mathbb{C}_{\delta}}$
the set of edges between a vertex of $\mathcal{V}_{\Omega_{\delta}}$
and a vertex of $\partial\mathcal{V}_{\Omega_{\delta}}$. The vertices
in $\partial\mathcal{V}_{\Omega_{\delta}}$ \emph{appear with multiplicity:}
if a vertex of $\mathcal{V}_{\mathbb{C}_{\delta}}\setminus\mathcal{V}_{\Omega_{\delta}}$
is at distance $\delta$ to several vertices of $\mathcal{V}_{\Omega_{\delta}}$,
then it appears as as many distinct elements of $\partial\mathcal{V}_{\Omega_{\delta}}$.
In other words, there is a one-to-one correspondence between $\partial\mathcal{V}_{\Omega_{\delta}}$
and $\partial\mathcal{E}_{\Omega_{\delta}}$.
\item We denote by $\partial\mathcal{V}_{\Omega_{\delta}^{*}}$ the centers
of the faces of $\mathbb{C}_{\delta}$ that are adjacent to a face
of $\Omega_{\delta}$. We denote by $\partial\mathcal{E}_{\Omega_{\delta}^{*}}\subset\partial\mathcal{E}_{\mathbb{C}_{\delta}^{*}}$
the set of dual edges between a vertex of $\mathcal{V}_{\Omega_{\delta}^{*}}$
and a vertex of $\partial\mathcal{V}_{\Omega_{\delta}^{*}}$. For
an edge $e\in\mathcal{E}_{\Omega_{\delta}}$ we denote by $e^{*}\in\mathcal{E}_{\Omega_{\delta}^{*}}$
its dual ($e$ and $e^{*}$ intersect at their midpoint).
\item We write $\overline{\mathcal{V}}_{\Omega_{\delta}}$ for $\mathcal{V}_{\Omega_{\delta}}\cup\partial\mathcal{V}_{\Omega_{\delta}}$
and $\overline{\mathcal{V}}_{\Omega_{\delta}^{*}}$ for $\mathcal{V}_{\Omega_{\delta}^{*}}\cup\partial\mathcal{V}_{\Omega_{\delta}^{*}}$.
\item We denote by $\mathcal{E}_{\Omega_{\delta}}^{h}\subset\mathcal{E}_{\Omega_{\delta}}$
the set of horizontal (i.e. parallel to the real axis) edges of $\Omega_{\delta}$
and by $\mathcal{E}_{\Omega_{\delta}}^{v}:=\mathcal{E}_{\Omega_{\delta}}\setminus\mathcal{E}_{\Omega_{\delta}}^{h}$
the set of the vertical ones.
\item We denote by $\Omega_{\delta}^{M}$ and the \emph{medial graph of
$\Omega_{\delta}$}: its vertices $\mathcal{V}_{\Omega_{\delta}^{M}}$
are the midpoints of the edges of $\mathcal{E}_{\Omega_{\delta}}\cup\partial\mathcal{E}_{\Omega_{\delta}}$
and the medial edges $\mathcal{E}_{\Omega_{\delta}^{M}}$ link midpoints
of incident edges of $\mathcal{E}_{\Omega_{\delta}}\cup\partial\mathcal{E}_{\Omega_{\delta}}$. 
\item We say that a family $\left(\Omega_{\delta}\right)_{\delta>0}$ of
discrete domains (with $\Omega_{\delta}\subset\mathbb{C}_{\delta}$
for each $\delta>0$) \emph{approximates} or \emph{discretizes }a
continuous domain $\Omega$ if for each $\delta>0$, $\Omega_{\delta}$
is the largest connected induced subgraph of $\mathbb{C}_{\delta}$
contained in $\Omega$. 
\end{itemize}

\subsubsection{Ising model with boundary conditions}

The Ising model (with \emph{free boundary condition}) on a finite
graph $\mathcal{G}$ (in this paper, $\mathcal{G}$ will be a discrete
domain $\Omega_{\delta}$ or its dual $\Omega_{\delta}^{*}$) at \emph{inverse
temperature} $\beta>0$ is a model whose state space $\Xi_{\mathcal{G}}$
is given by $\Xi_{\mathcal{G}}:=\left\{ \left(\sigma_{x}\right)_{x\in\mathcal{V}_{\mathcal{G}}}:\sigma_{x}\in\left\{ \pm1\right\} \right\} $:
a state assigns to every vertex $x$ of $\mathcal{G}$ a \emph{spin}
$\sigma_{x}\in\left\{ \pm1\right\} $. The probability of a configuration
$\sigma\in\Xi_{\mathcal{G}}$ is
\[
\mathbb{P}_{\mathcal{G}}^{\beta,\mathrm{free}}\left\{ \sigma\right\} \,:=\,\frac{1}{\mathcal{Z}_{\mathcal{G}}^{\beta,\mathrm{free}}}e^{-\beta H_{\mathcal{G}}^{\beta,\mathrm{free}}\left(\sigma\right)},
\]
with the \emph{energy} (or Hamiltonian) $H_{\mathcal{G}}^{\beta,\mathrm{free}}$
of a configuration $\sigma$ given by 
\[
H_{\mathcal{G}}^{\beta,\mathrm{free}}\left(\sigma\right)\,:=\,-\sum_{\left\langle x,y\right\rangle \in\mathcal{E}_{\mathcal{G}}}\sigma_{x}\sigma_{y},
\]
and the \emph{partition function} $\mathcal{Z}_{\mathcal{G}}^{\beta,\mathrm{free}}$
by 
\[
\mathcal{Z}_{\mathcal{G}}^{\beta,\mathrm{free}}\,:=\,\sum_{\sigma\in\Xi_{\mathcal{G}}}e^{-\beta H\left(\sigma\right)}.
\]

Given a graph $\mathcal{G}$ with boundary vertices $\partial\mathcal{V}_{\mathcal{G}}$
(like $\Omega_{\delta}^{*}$ with $\partial\mathcal{V}_{\Omega_{\delta}^{*}}$)
the Ising model on $\mathcal{G}$ with\emph{ $+$ boundary }condition\emph{
}is defined as the Ising model on $\mathcal{G}$, with extra spins
located at the vertices of $\partial\mathcal{V}_{\mathcal{G}}$ that
are set to $+1$ and with energy 
\[
H_{\mathcal{G}}^{\beta,+}\left(\sigma\right)\,:=\,-\sum_{\left\langle x,y\right\rangle \in\mathcal{\overline{E}_{\mathcal{G}}}}\sigma_{x}\sigma_{y},
\]
where $\overline{\mathcal{E}}_{\mathcal{G}}$ is the set of edges
linking vertices of $\mathcal{V}_{\mathcal{G}}\cup\partial\mathcal{V}_{\mathcal{G}}$. 

In this paper, we will be interested in the Ising model with free
and $+$ boundary conditions on discrete square grid domains $\Omega_{\delta}$
at the critical inverse temperature $\beta_{c}:=\frac{1}{2}\ln\left(\sqrt{2}+1\right)$,
when the mesh size $\delta$ is small. 

We will from now on omit the inverse temperature parameter $\beta$
in the notation and will denote by $\mathbb{P}_{\mathcal{G}}^{\mathrm{free}}$
and $\mathbb{P}_{\mathcal{G}}^{+}$ the probability measures of the
Ising model on $\mathcal{G}$ at $\beta=\beta_{c}$ with free and
$+$ boundary conditions and by $\mathbb{E}_{\mathcal{G}}^{\mathrm{free}}$
and $\mathbb{E}_{\mathcal{G}}^{+}$ the corresponding expectations.

\subsection{\label{sub:the-energy-density}The energy density}

\begin{figure}
\includegraphics[scale=0.3]{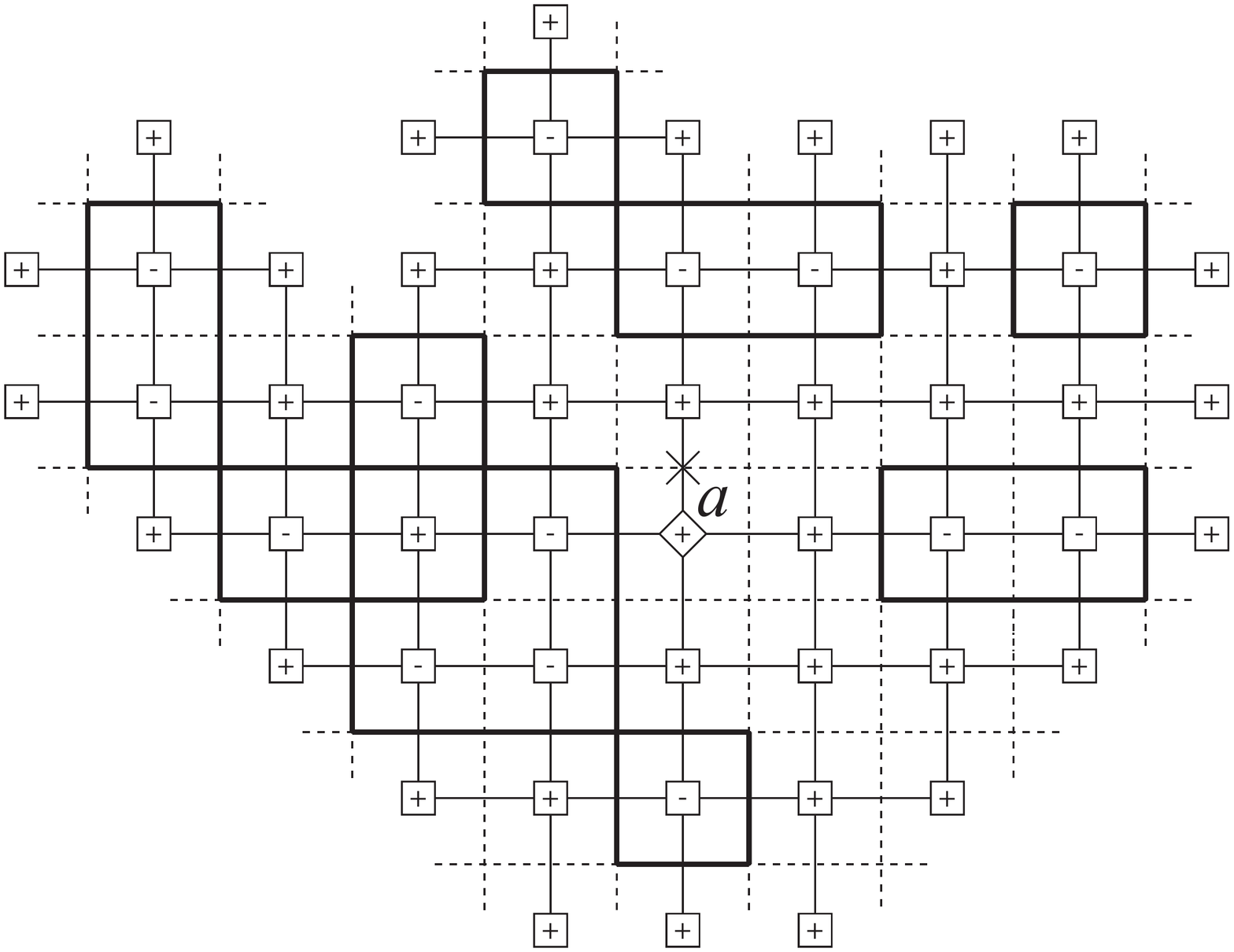}\caption{\label{fig:spins-lt}The Ising model on $\mathcal{V}_{\Omega_{\delta}^{*}}$
with $+$ boundary condition, with the contours corresponding to its
low-temperature expansion.}
\end{figure}

\begin{figure}
\includegraphics[scale=0.3]{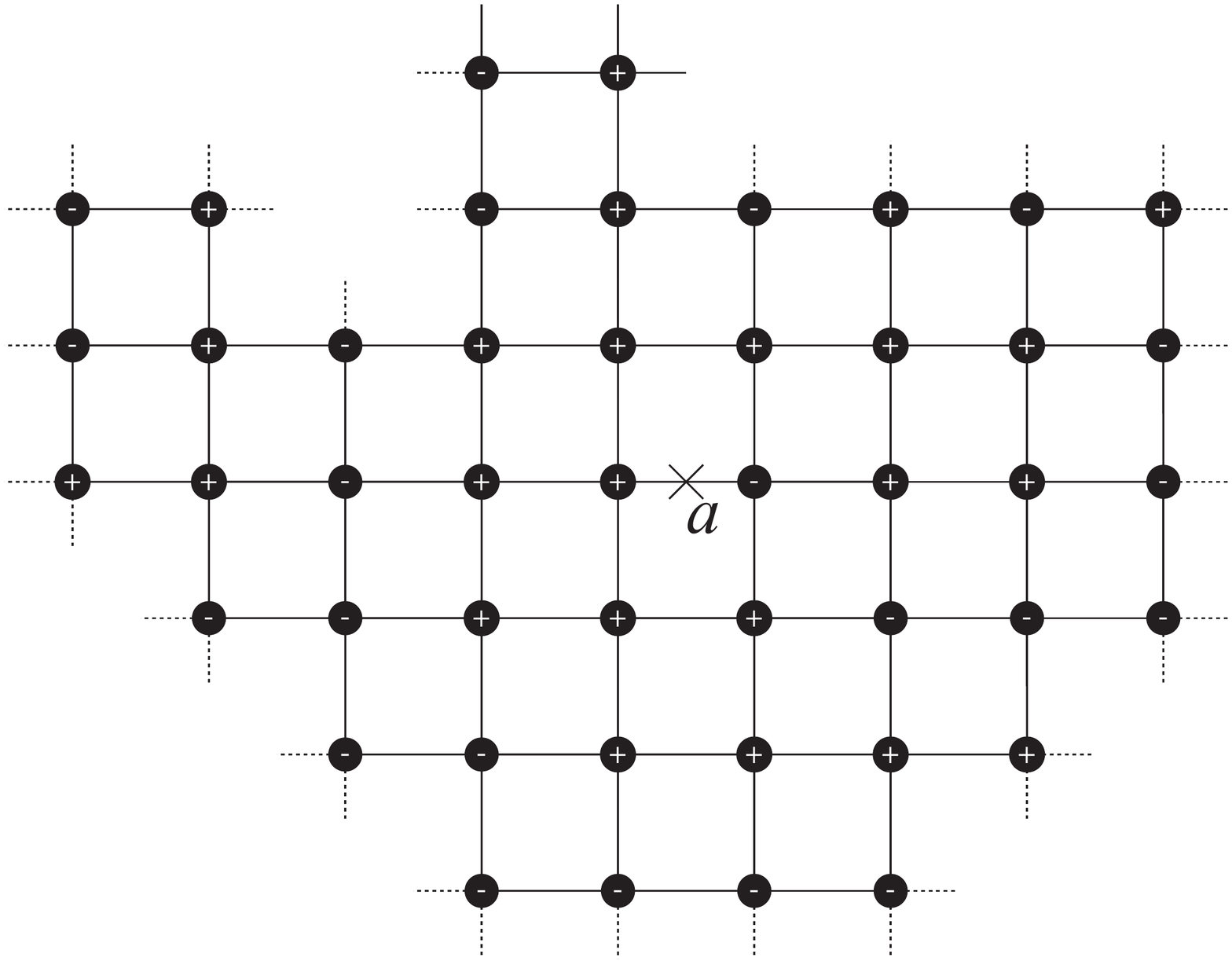}\caption{\label{fig:spins-ht}The Ising model on $\mathcal{V}_{\Omega_{\delta}}$
with free boundary condition.}
\end{figure}

Let $\Omega_{\delta}$ be a discrete domain and let $a_{\delta}\in\mathcal{V}_{\Omega_{\delta}^{M}}$
be the midpoint of a horizontal edge of $\Omega_{\delta}$. We introduce
the two quantities $\left\langle \epsilon_{\delta}\left(a_{\delta}\right)\right\rangle _{\Omega_{\delta}}^{\mathrm{free}}$
and $\left\langle \epsilon_{\delta}\left(a_{\delta}\right)\right\rangle _{\Omega_{\delta}^{*}}^{+}$,
called \emph{average energy density} (with free and $+$ boundary
conditions), defined by 
\begin{eqnarray*}
\left\langle \epsilon_{\delta}\left(a_{\delta}\right)\right\rangle _{\Omega_{\delta}}^{\mathrm{free}} & \,:=\, & \mathbb{E}_{\Omega_{\delta}}^{\mathrm{free}}\left[\sigma_{e_{\delta}}\sigma_{w_{\delta}}-\frac{\sqrt{2}}{2}\right],\\
\left\langle \epsilon_{\delta}\left(a_{\delta}\right)\right\rangle _{\Omega_{\delta}^{*}}^{+} & \,:=\, & \mathbb{E}_{\Omega_{\delta}^{*}}^{+}\left[\sigma_{n_{\delta}}\sigma_{s_{\delta}}-\frac{\sqrt{2}}{2}\right],
\end{eqnarray*}
 where $\left\langle e_{\delta},w_{\delta}\right\rangle \in\mathcal{E}_{\Omega_{\delta}}$
and $\left\langle n_{\delta},s_{\delta}\right\rangle \in\mathcal{E}_{\Omega_{\delta}^{*}}$
are respectively the (horizontal) edge and the dual (vertical) edge,
the midpoint of both of which is $a_{\delta}$ (see Figures \ref{fig:spins-lt}
and \ref{fig:spins-ht}). The quantity $\frac{\sqrt{2}}{2}$ is the
infinite-volume limit of the product of two adjacent spins (it can
be found in \cite{mccoy-wu}, Chapter VIII, Formula 4.12, for instance).
The energy density field is the fluctuation of the product of adjacent
spins around this limit: it measures the distribution of the energy
\emph{$H$} among the edges, as a function of their locations. We
are considering horizontal edges on $\Omega_{\delta}$ and vertical
edges on $\Omega_{\delta}^{*}$ for concreteness and for making the
notation simpler, but our results are rotationally invariant.

We can now state the main result of this paper, which is the conformal
covariance of the average energy density:
\begin{thm}
\label{thm:main-thm}Let $\Omega$ be a $\mathcal{C}^{1}$ simply
connected domain and let $a\in\Omega$. Consider a family $\left(\Omega_{\delta}\right)_{\delta>0}$
of discrete domains approximating $\Omega$ and for each $\delta>0$,
let $a_{\delta}\in\mathcal{V}_{\Omega_{\delta}^{M}}$ be the midpoint
of horizontal edge that is the closest to $a$. Then as $\delta\to0$,
uniformly on the compact subsets of $\Omega$, we have
\begin{eqnarray*}
\frac{1}{\delta}\left\langle \epsilon_{\delta}\left(a_{\delta}\right)\right\rangle _{\Omega_{\delta}^{*}}^{+} & \to & \phantom{-}\frac{1}{2\pi}\ell_{\Omega}\left(a\right),\\
\frac{1}{\delta}\left\langle \epsilon_{\delta}\left(a_{\delta}\right)\right\rangle _{\Omega_{\delta}^{*}}^{\mathrm{free}} & \to & -\frac{1}{2\pi}\ell_{\Omega}\left(a\right),
\end{eqnarray*}
$\ell_{\Omega}\left(a\right)$ being the hyperbolic metric element
of $\Omega$ at $a$. Namely, $\ell_{\Omega}\left(a\right):=2\psi_{a}'\left(a\right)$,
where $\psi_{a}$ is the conformal mapping from $\Omega$ to the unit
disk $\mathbb{D}:=\left\{ z\in\mathbb{C}:\left|z\right|<1\right\} $
such that $\psi_{a}\left(a\right)=$0 and $\psi_{a}'\left(a\right)>0$. 
\end{thm}
The proof will be given in Section \ref{sub:convergence-results}
\begin{cor}
The conclusions of Theorem \ref{thm:main-thm} hold under the assumption
that $\Omega$ is a Jordan domain.\end{cor}
\begin{proof}
We have that $\left\langle \epsilon_{\delta}\left(a_{\delta}\right)\right\rangle _{\Omega_{\delta}}^{+}$
and $\left\langle \epsilon_{\delta}\left(a_{\delta}\right)\right\rangle _{\Omega_{\delta}}^{\mathrm{free}}$
are monotone respectively non-increasing and non-decreasing with respect
to the discrete domain $\Omega_{\delta}$, as follows easily from
the FKG inequality applied to the FK representation of the model (see
\cite{grimmett} Chapters 1 and 2, for instance): if $\Omega_{\delta}\subset\tilde{\Omega}_{\delta}$,
we have
\begin{eqnarray*}
\left\langle \epsilon_{\delta}\left(a_{\delta}\right)\right\rangle _{\Omega_{\delta}^{*}}^{+} & \geq & \left\langle \epsilon_{\delta}\left(a_{\delta}\right)\right\rangle _{\tilde{\Omega}_{\delta}^{*}}^{+},\\
\left\langle \epsilon_{\delta}\left(a_{\delta}\right)\right\rangle _{\Omega_{\delta}}^{\mathrm{free}} & \leq & \left\langle \epsilon_{\delta}\left(a_{\delta}\right)\right\rangle _{\tilde{\Omega}_{\delta}}^{\mathrm{free}}.
\end{eqnarray*}
If $\Omega$ is a Jordan domain, we can approximate $\Omega$ by monotone
increasing and decreasing sequences of smooth domains, for which Theorem
\ref{thm:main-thm} applies, and deduce the result for $\Omega$.
\end{proof}
The central idea for proving Theorem \ref{thm:main-thm} is to introduce
a discrete fermionic spinor which is a two-point function $f_{\Omega_{\delta}}\left(a,z\right)$;
it is defined in the next subsection. We then relate $f_{\Omega_{\delta}}$
to the average energy density and prove its convergence to a holomorphic
function $f_{\Omega}$.

\subsection{\label{sub:contour-stat-disc-hol-spin}Contour statistics and discrete
fermionic spinor}

\subsubsection{Contour statistics}

Let $\Omega_{\delta}$ be a discrete domain. We denote by $\mathcal{C}_{\Omega_{\delta}}$
the set of edge collections $\omega\subset\mathcal{E}_{\Omega_{\delta}}$
such that every vertex $v\in\mathcal{V}_{\Omega_{\delta}}$ belongs
to an even number of edges of $\omega$: in other words, by Euler's
theorem for walks, the edge collections $\omega\in\mathcal{C}_{\Omega_{\delta}}$
are the ones that consist of edges forming (not necessarily simple)
closed contours. For an edge $e\in\mathcal{E}_{\Omega_{\delta}}$,
we denote by $\mathcal{C}_{\Omega_{\delta}}^{\left\{ e+\right\} }$
the set of configurations $\omega\in\mathcal{C}_{\Omega_{\delta}}$
that \emph{do not contain} $e$ and by $\mathcal{C}_{\Omega_{\delta}}^{\left\{ e-\right\} }$
the set of configurations that \emph{do contain} $e$.

Set $\alpha:=\sqrt{2}-1$. For a collection of edges $\omega\subset\mathcal{E}_{\Omega_{\delta}}$,
we denote by $\left|\omega\right|$ its cardinality. For $e\in\mathcal{E}_{\Omega_{\delta}}$
we define
\begin{eqnarray*}
\mathbf{Z}_{\Omega_{\delta}} & \,:=\, & \sum_{\omega\in\mathcal{C}_{\Omega_{\delta}}}\alpha^{\left|\omega\right|},\\
\mathbf{Z}_{\Omega_{\delta}}^{\left\{ e+\right\} } & \,:=\, & \sum_{\omega\in\mathcal{C}_{\Omega_{\delta}}^{\left\{ e+\right\} }}\alpha^{\left|\omega\right|},\\
\mathbf{Z}_{\Omega_{\delta}}^{\left\{ e-\right\} } & \,:=\, & \sum_{\omega\in\mathcal{C}_{\Omega_{\delta}}^{\left\{ e-\right\} }}\alpha^{\left|\omega\right|}.
\end{eqnarray*}
We now have the following representation of the energy density in
terms of contour statistics. 

\begin{prop}
\label{pro:contour-statistics}Let $e\in\mathcal{E}_{\Omega_{\delta}}^{h}$
be a horizontal edge and let its midpoint be $a\in\mathcal{V}_{\Omega_{\delta}^{M}}$.
Then we have 
\begin{eqnarray*}
\left\langle \epsilon_{\delta}\left(a\right)\right\rangle _{\Omega_{\delta}^{*}}^{+} & = & \frac{\mathbf{Z}_{\Omega_{\delta}}^{\left\{ e+\right\} }}{\mathbf{Z}_{\Omega_{\delta}}}-\frac{\mathbf{Z}_{\Omega_{\delta}}^{\left\{ e-\right\} }}{\mathbf{Z}_{\Omega_{\delta}}}-\frac{\sqrt{2}}{2}=2\frac{\mathbf{Z}_{\Omega_{\delta}}^{\left\{ e+\right\} }}{\mathbf{Z}_{\Omega_{\delta}}}-\frac{\sqrt{2}+2}{2},\\
\left\langle \epsilon_{\delta}\left(a\right)\right\rangle _{\Omega_{\delta}}^{\mathrm{free}} & = & -\left\langle \epsilon_{\delta}\left(a\right)\right\rangle _{\Omega_{\delta}^{*}}^{+}.
\end{eqnarray*}
\end{prop}
\begin{proof}
From the low-temperature expansion of the Ising model (see \cite{palmer},
Chapter 1, for instance), there is a natural bijection between the
configurations of spins $\sigma$ on $\mathcal{V}_{\Omega_{\delta}^{*}}$
with $+$ boundary condition on $\partial\mathcal{V}_{\Omega_{\delta}^{*}}$
and the edges collections $\omega\in\mathcal{C}_{\Omega_{\delta}}$:
one puts an edge $\mathfrak{e}\in\mathcal{E}_{\Omega_{\delta}}$ in
the edge collection $\omega$ if the spins of $\sigma$ at the endpoints
of the dual edge $\mathfrak{e}^{*}\in\mathcal{E}_{\Omega_{\delta}^{*}}$
are different. It is easy to see that the probability measure on $\mathcal{C}_{\Omega_{\delta}}$
induced by this bijection gives to each edge collection $\omega\in\mathcal{C}_{\Omega_{\delta}}$
a weight proportional to $\left(e^{-2\beta}\right)^{\left|\omega\right|}$,
hence to $\alpha^{\left|\omega\right|}$, where $\alpha=\sqrt{2}-1$
as above (since $\beta=\beta_{c}=\frac{1}{2}\ln\left(\sqrt{2}+1\right)$).
The event that the spins at two adjacent dual vertices $x,y\in\mathcal{V}_{\Omega_{\delta}^{*}}$
are the same (respectively are different) corresponds through the
natural bijection to $\mathcal{C}_{\Omega_{\delta}}^{\left\{ e^{+}\right\} }$
(respectively $\mathcal{C}_{\Omega_{\delta}}^{\left\{ e^{-}\right\} }$),
where $e\in\mathcal{E}_{\Omega_{\delta}}$ is such that $e^{*}=\left\langle x,y\right\rangle $.
Using that $\mathbf{Z}_{\Omega_{\delta}}^{\left\{ e+\right\} }+\mathbf{Z}_{\Omega_{\delta}}^{\left\{ e-\right\} }=\mathbf{Z}_{\Omega_{\delta}}$,
we deduce the first identity.

From the so-called high-temperature expansion (see \cite{palmer},
Chapter 1) we have that for the Ising model on $\Omega_{\delta}$
with free boundary condition, the correlation of two spins $z_{1},z_{2}\in\mathcal{V}_{\Omega_{\delta}}$
is equal to 
\[
\frac{\sum_{\tilde{\omega}\in\mathcal{C}_{\Omega_{\delta}}\left(z_{1},z_{2}\right)}\left(\tanh\left(\beta\right)\right)^{\left|\tilde{\omega}\right|}}{\sum_{\omega\in\mathcal{C}_{\Omega_{\delta}}}\left(\tanh\left(\beta\right)\right)^{\left|\omega\right|}},
\]
where $\mathcal{C}_{\Omega_{\delta}}\left(z_{1},z_{2}\right)$ is
the set of edge collections $\tilde{\omega}$ such that every vertex
in $\mathcal{V}_{\Omega_{\delta}}\setminus\left\{ z_{1},z_{2}\right\} $
belongs to an even number of edges of $\tilde{\omega}$ and such that
$z_{1},z_{2}$ both belong to an odd number of edges of $\tilde{\omega}$.
At $\beta=\beta_{c}$, we have $\tanh\left(\beta\right)=\alpha$ (the
fact that $\tanh\beta_{c}=e^{-2\beta_{c}}$ actually characterizes
$\beta_{c}$). 

Let us now take $z_{1},z_{2}$ adjacent, set $e:=\left\langle z_{1},z_{2}\right\rangle \in\mathcal{E}_{\Omega_{\delta}}$,
and denote by $\mathcal{C}_{\Omega_{\delta}}^{+}\left(z_{1},z_{2}\right)$
and $\mathcal{C}_{\Omega_{\delta}}^{-}\left(z_{1},z_{2}\right)$ the
sets of $\tilde{\omega}\in\mathcal{C}_{\Omega_{\delta}}\left(z_{1},z_{2}\right)$
such that $e\in\tilde{\omega}$ and $e\notin\tilde{\omega}$ respectively.
From each $\tilde{\omega}\in\mathcal{C}_{\Omega_{\delta}}^{+}\left(z_{1},z_{2}\right)$,
we can remove $e$ and obtain an edge collection in $\mathcal{C}_{\Omega_{\delta}}^{\left\{ e+\right\} }$
(this map $\mathcal{C}_{\Omega_{\delta}}^{+}\left(z_{1},z_{2}\right)\to\mathcal{C}_{\Omega_{\delta}}^{\left\{ e+\right\} }$
is bijective) and to each $\tilde{\omega}\in\mathcal{C}_{\Omega_{\delta}}^{-}\left(z_{1},z_{2}\right)$,
we can add $e$ and obtain an edge collection in $\mathcal{C}_{\Omega_{\delta}}^{\left\{ e-\right\} }$
. Hence we have
\begin{eqnarray*}
\left\langle \epsilon_{\delta}\left(a\right)\right\rangle _{\Omega_{\delta}}^{\mathrm{free}} & = & \frac{\sum_{\tilde{\omega}\in\mathcal{C}_{\Omega_{\delta}}^{+}\left(z_{1},z_{2}\right)}\alpha^{\left|\tilde{\omega}\right|}}{\mathbf{Z}_{\Omega_{\delta}}}+\frac{\sum_{\tilde{\omega}\in\mathcal{C}_{\Omega_{\delta}}^{-}\left(z_{1},z_{2}\right)}\alpha^{\left|\tilde{\omega}\right|}}{\mathbf{Z}_{\Omega_{\delta}}}-\frac{\sqrt{2}}{2}\\
 & = & \frac{\alpha\sum_{\omega\in\mathcal{C}_{\Omega_{\delta}}^{\left\{ e+\right\} }}\alpha^{\left|\omega\right|}}{\mathbf{Z}_{\Omega_{\delta}}}+\frac{\alpha^{-1}\sum_{\omega\in\mathcal{C}_{\Omega_{\delta}}^{\left\{ e-\right\} }}\alpha^{\left|\omega\right|}}{\mathbf{Z}_{\Omega_{\delta}}}-\frac{\sqrt{2}}{2}\\
 & = & \frac{\alpha\mathbf{Z}_{\Omega_{\delta}}^{\left\{ e+\right\} }+\alpha^{-1}\mathbf{Z}_{\Omega_{\delta}}^{\left\{ e-\right\} }}{\mathbf{Z}_{\Omega_{\delta}}}-\frac{\sqrt{2}}{2}.
\end{eqnarray*}
Using the relation $\mathbf{Z}_{\Omega_{\delta}}^{\left\{ e+\right\} }+\mathbf{Z}_{\Omega_{\delta}}^{\left\{ e-\right\} }=\mathbf{Z}_{\Omega_{\delta}}$,
we obtain the second identity. 
\end{proof}

\subsubsection{Discrete fermionic  spinor in bounded domain}

\begin{figure}
\includegraphics[scale=0.4]{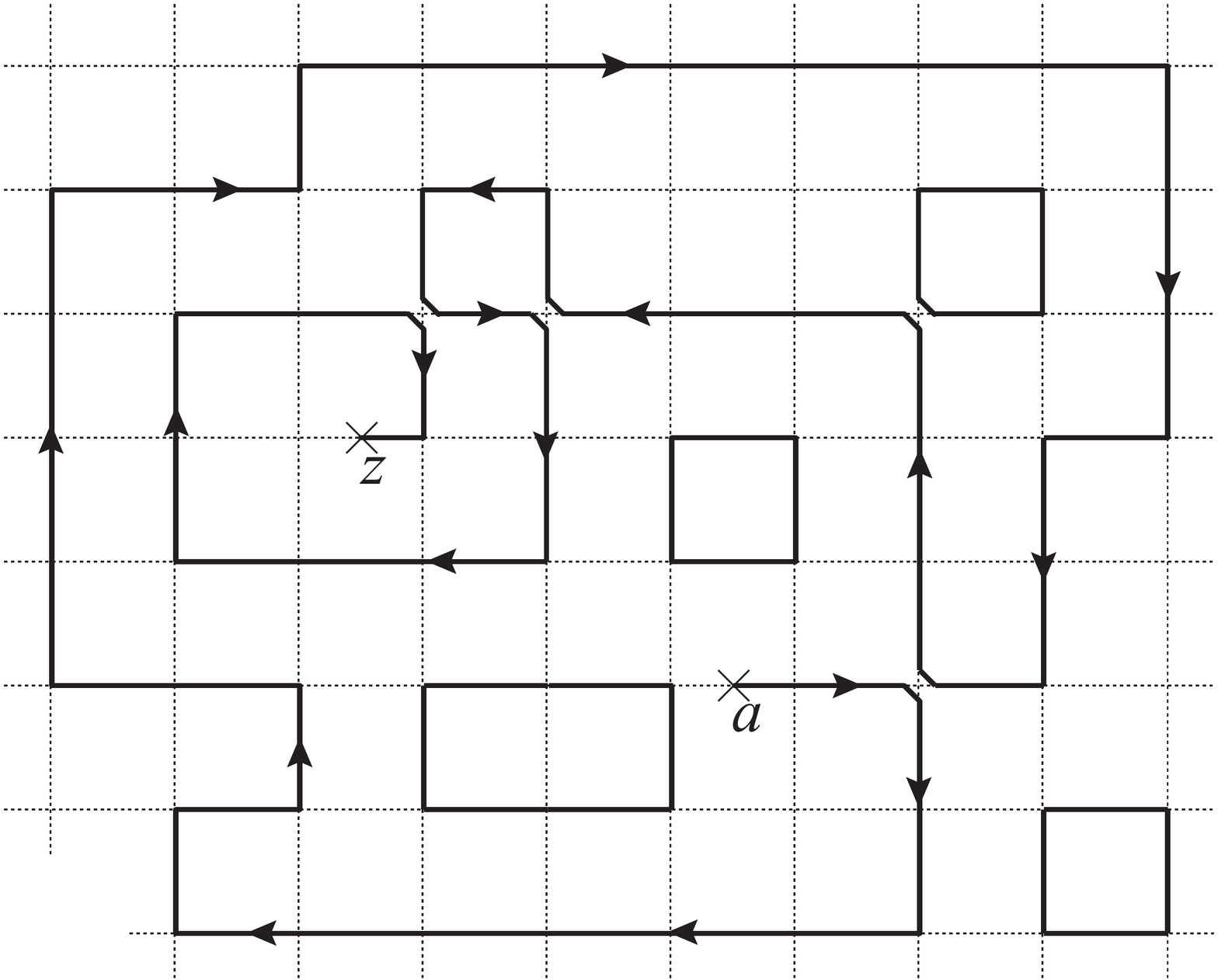}\caption{\label{fig:configuration-walk}An admissible walk in a configuration
in $\mathcal{C}_{\Omega_{\delta}}\left(a,z\right)$. There are sixteen
different choices of such walks in this configuration. }
\end{figure}

If $a\in\mathcal{V}_{\Omega_{\delta}^{M}}$ is the midpoint of a horizontal
edge $e_{1}\in\mathcal{E}_{\Omega_{\delta}}^{h}$ and $z\in\mathcal{V}_{\Omega_{\delta}^{M}}$
is the midpoint of an arbitrary edge $e_{2}\in\mathcal{E}_{\Omega_{\delta}}\cup\partial\mathcal{E}_{\Omega_{\delta}}$,
we denote by $\mathcal{C}_{\Omega_{\delta}}\left(a,z\right)$ the
set of $\gamma$ consisting of edges of $\mathcal{E}_{\Omega_{\delta}}\setminus\left\{ e_{1},e_{2}\right\} $
and of two \emph{half-edges} (half of an edge between its midpoint
and one of its ends) such that
\begin{itemize}
\item one of the half-edges has endpoints $a,a+\frac{\delta}{2}$;
\item the other half-edge is incident to $z$;
\item every vertex $v\in\mathcal{V}_{\Omega_{\delta}}$ belongs to an even
number of edges or half-edges of $\gamma$.
\end{itemize}
For a configuration $\gamma\in\mathcal{C}_{\Omega_{\delta}}\left(a,z\right)$,
we call \emph{admissible walk along $\gamma$}, a sequence $e_{0},e_{1},\ldots,e_{n}$,
such that 
\begin{itemize}
\item $e_{0}$ is the half-edge incident to $a$;
\item $e_{n}$ is the half-edge incident to $z$;
\item $e_{1},\ldots,e_{n-1}\in\mathcal{E}_{\Omega_{\delta}}$ are edges;
\item $e_{j}$ and $e_{j+1}$ are incident for each $j\in\left\{ 0,\ldots,n-1\right\} $;
\item each edge appears at most once in the walk;
\item when one follows the walk and arrives at a vertex that belongs to
four edges or half-edges of $\gamma$ (we call this an \emph{ambiguity}),
one either turns left or right (going straight in that case is forbidden).
\end{itemize}
It is easy to see that, for any $\gamma\in\mathcal{C}_{\Omega_{\delta}}\left(a,z\right)$,
such a walk always exists, though in general it is not unique (see
Figure \ref{fig:configuration-walk}).

Given a configuration $\gamma\in\mathcal{C}_{\Omega_{\delta}}\left(a,z\right)$
and an admissible walk along $\gamma$, we define the \emph{winding
number} $\mathbf{W}$ of $\gamma$, denoted $\mathbf{W}\left(\gamma\right)\in\mathbb{R}/4\pi\mathbb{Z}$,
by 
\[
\mathbf{W}\left(\gamma\right)\,:=\,\frac{\pi}{2}\left(n_{\ell}-n_{r}\right),
\]
where $n_{\ell}$ and $n_{r}$ are respectively the numbers of left
turns and right turns that the admissible walk makes from $a$ to
$z$: it is the total rotation of the walk between $a$ and $z$,
measured in radians. More generally, we define the winding number
of a rectifiable curve as its total rotation from its initial point
to its final point, measured in radians. The following lemma shows
that the winding number (modulo $4\pi$) of a configuration $\gamma\in\mathcal{C}_{\Omega_{\delta}}\left(a,z\right)$
is actually independent of the choice of the walk on $\gamma$.

\begin{lem}
\label{lem:well-definedness}For any $\gamma\in\mathcal{C}_{\Omega_{\delta}}\left(a,z\right)$,
the winding number $\mathbf{W}\left(\gamma\right)\in\mathbb{R}/4\pi\mathbb{Z}$
is independent of the choice of an admissible walk along $\gamma$.
\end{lem}
The proof is given in Appendix A.

Thanks to Lemma \ref{lem:well-definedness}, we can now define the
discrete fermionic spinor $f_{\Omega_{\delta}}$ that will be instrumental
in our studies of the energy density.
\begin{defn}
\label{def:disc-hol-spin-bded-domain}For any midpoint of a horizontal
edge $a\in\mathcal{V}_{\Omega_{\delta}^{M}}$, we define the discrete
fermionic spinor $f_{\Omega_{\delta}}\left(a,\cdot\right):\mathcal{V}_{\Omega_{\delta}^{M}}\to\mathbb{C}$
by
\begin{eqnarray*}
f_{\Omega_{\delta}}\left(a,z\right) & \,:=\, & \frac{1}{\mathbf{Z}_{\Omega_{\delta}}}\sum_{\gamma\in\mathcal{C}_{\Omega_{\delta}}\left(a,z\right)}\alpha^{\left|\gamma\right|}e^{-\frac{i}{2}\mathbf{W}\left(\gamma\right)},\\
f_{\Omega_{\delta}}\left(a,a\right) & \,:=\, & \frac{\mathbf{Z}_{\Omega_{\delta}}^{\left\{ e+\right\} }}{\mathbf{Z}_{\Omega_{\delta}}},
\end{eqnarray*}
where $\left|\gamma\right|$ denotes the number of edges and half-edges
of $\gamma$, with the half-edges contributing $\frac{1}{2}$ each.
\end{defn}
In this way, $z\mapsto f_{\Omega_{\delta}}\left(a,z\right)$ is a
function, whose value at $z=a$ gives (up to an additive constant)
the average energy density at $a$. As we will see, moving the point
$z$ accross the domain will allow to gain information about the effect
of the geometry of the domain on the energy density.

\subsubsection{\label{sub:disc-hol-spin-full-plane}Discrete fermionic spinor in
the full plane}

As mentioned above, the $\frac{\sqrt{2}}{2}$ appearing in the definition
of the average energy density (Section \ref{sub:the-energy-density})
is the infinite volume limit (or full-plane) average product of two
adjacent spins, which one has to subtract in order for the effect
of the shape of the domain to be studied. We now introduce a full-plane
version of the discrete fermionic spinor, whose definition a priori
seems quite different from the bounded domain version. It will allow
to represent the energy density (with the correct additive constant)
in terms of the difference of two discrete fermionic spinors.
\begin{defn}
\label{def:full-plane-spinor}For $a,z\in\mathcal{V}_{\Omega_{\delta}^{M}}$
with $a\neq z$ and $a$ being the midpoint of an horizontal edge,
define $f_{\mathbb{C}_{\delta}}$ by
\begin{eqnarray*}
f_{\mathbb{C}_{\delta}}\left(a,z\right) & \,:=\, & 2\cos\left(\frac{\pi}{8}\right)e^{\frac{\pi i}{8}}\left(C_{0}\left(\frac{2\left(a+\frac{\delta}{2}\right)}{\delta},\frac{2z}{\delta}\right)+C_{0}\left(\frac{2\left(a-\frac{i\delta}{2}\right)}{\delta},\frac{2z}{\delta}\right)\right)\\
 &  & +2\sin\left(\frac{\pi}{8}\right)e^{-\frac{3\pi i}{8}}\left(C_{0}\left(\frac{2\left(a-\frac{\delta}{2}\right)}{\delta},\frac{2z}{\delta}\right)+C_{0}\left(\frac{2\left(a+\frac{i\delta}{2}\right)}{\delta},\frac{2z}{\delta}\right)\right),
\end{eqnarray*}
where $C_{0}\left(z_{1},z_{2}\right):=C_{0}\left(0,z_{2}-z_{1}\right)$
is the dimer coupling function of Kenyon (see \cite{kenyon-i}), defined
by
\[
C\left(0,x+iy\right):=\frac{1}{4\pi^{2}}\int_{0}^{2\pi}\int_{0}^{2\pi}\frac{e^{i\left(x\theta-y\phi\right)}}{2i\sin\left(\theta\right)+2\sin\left(\phi\right)}d\theta d\phi.
\]
We set $f_{\mathbb{C}_{\delta}}\left(a,a\right):=\frac{2+\sqrt{2}}{4}$. 
\end{defn}
From our definitions up to now and from Proposition \ref{pro:contour-statistics},
we deduce the following:
\begin{lem}
\label{lem:representation-ed-observable}Let $\Omega_{\delta}$ be
a discrete domain and $a\in\mathcal{V}_{\Omega_{\delta}^{M}}$ be
the midpoint of a horizontal edge of $\Omega_{\delta}$. Then the
average energy density can be represented as 
\begin{eqnarray*}
\left\langle \epsilon_{\delta}\left(a\right)\right\rangle _{\Omega_{\delta}^{*}}^{+} & = & \phantom{-}2\left(f_{\Omega_{\delta}}-f_{\mathbb{C}_{\delta}}\right)\left(a,a\right),\\
\left\langle \epsilon_{\delta}\left(a\right)\right\rangle _{\Omega_{\delta}}^{\mathrm{free}} & = & -2\left(f_{\Omega_{\delta}}-f_{\mathbb{C}_{\delta}}\right)\left(a,a\right).
\end{eqnarray*}

\end{lem}

\subsection{\label{sub:convergence-results}Convergence results and proof of
Theorem \ref{thm:main-thm}}

The core of this paper is the convergence of the discrete fermionic
spinors to continuous ones, which are holomorphic functions. Let us
define these functions first: for $a,z\in\Omega$ with $a\neq z$,
we define
\begin{eqnarray*}
f_{\Omega}\left(a,z\right) & := & \frac{1}{2\pi}\sqrt{\psi_{a}'\left(a\right)}\sqrt{\psi_{a}'\left(z\right)}\frac{\psi_{a}\left(z\right)+1}{\psi_{a}\left(z\right)},\\
f_{\mathbb{C}}\left(a,z\right) & := & \frac{1}{2\pi\left(z-a\right)},
\end{eqnarray*}
where $\psi_{a}$ is the unique conformal mapping from $\Omega$ to
the unit disk $\mathbb{D}$ with $\psi_{a}\left(a\right)=0$ and $\psi_{a}'\left(a\right)>0$
(this mapping exists by the Riemann mapping theorem). Note that $z\mapsto f_{\Omega}\left(a,z\right)$
and $z\mapsto f_{\mathbb{C}}\left(a,z\right)$ both have a simple
pole of residue $\frac{1}{2\pi}$ at $z=a$ and that $\left(f_{\Omega}-f_{\mathbb{C}}\right)\left(a,z\right)$
hence extends holomorphically to $z=a$.

We can now state the key theorem of this paper:
\begin{thm}
\label{thm:key-thm}For each $\delta>0$, identify $a\in\mathbb{C}$
with the closest midpoint of a horizontal edge of $\mathbb{C}_{\delta}$
and $z\in\mathbb{C}$ with the closest midpoint of an edge of $\mathbb{C}_{\delta}$.
Then we have the following convergence results
\begin{eqnarray*}
\frac{1}{\delta}f_{\Omega_{\delta}}\left(a,z\right) & \underset{\delta\to0}{\longrightarrow} & f_{\Omega}\left(a,z\right)\,\,\,\,\forall a,z\in\Omega:a\neq z,\\
\frac{1}{\delta}f_{\mathbb{C}_{\delta}}\left(a,z\right) & \underset{\delta\to0}{\longrightarrow} & f_{\mathbb{C}}\left(a,z\right)\,\,\,\,\forall a,z\in\mathbb{C}:a\neq z,\\
\frac{1}{\delta}\left(f_{\Omega_{\delta}}-f_{\mathbb{C}_{\delta}}\right)\left(a,z\right) & \underset{\delta\to0}{\longrightarrow} & \left(f_{\Omega}-f_{\mathbb{C}}\right)\left(a,z\right)\,\,\,\,\forall a,z\in\Omega.
\end{eqnarray*}
where the convergence of $\frac{1}{\delta}f_{\Omega_{\delta}}$ is
uniform on the compact subsets of $\Omega\times\Omega$ away from
the diagonal $\left\{ \left(w,w\right):w\in\Omega\right\} $, the
convergence of $\frac{1}{\delta}f_{\mathbb{C}_{\delta}}$ is uniform
on $\mathbb{C}\times\mathbb{C}$ away from the diagonal $\left\{ \left(w,w\right):w\in\mathbb{C}\right\} $
and the convergence of $\frac{1}{\delta}\left(f_{\Omega_{\delta}}-f_{\mathbb{C}_{\delta}}\right)$
is uniform on the compact subsets of $\Omega\times\Omega$.
\end{thm}
From this result, the proof of the main theorem follows readily: since
we have (Lemma \ref{lem:representation-ed-observable}) 
\begin{eqnarray*}
\left\langle \epsilon_{\delta}\left(a\right)\right\rangle _{\Omega_{\delta}^{*}}^{+} & = & \phantom{-}2\left(f_{\Omega_{\delta}}-f_{\mathbb{C}_{\delta}}\right)\left(a,a\right),\\
\left\langle \epsilon_{\delta}\left(a\right)\right\rangle _{\Omega_{\delta}}^{\mathrm{free}} & = & -2\left(f_{\Omega_{\delta}}-f_{\mathbb{C}_{\delta}}\right)\left(a,a\right),
\end{eqnarray*}
and since $\frac{1}{\delta}\left(f_{\Omega_{\delta}}-f_{\mathbb{C}_{\delta}}\right)$
converges to $\left(f_{\Omega}-f_{\mathbb{C}}\right)\left(a,a\right)$,
it suffices to check that
\[
\left(\sqrt{\psi_{a}'\left(a\right)}\sqrt{\psi_{a}'\left(z\right)}\frac{\psi_{a}\left(z\right)+1}{\psi_{a}\left(z\right)}-\frac{1}{z-a}\right)\underset{z\to a}{\longrightarrow}\psi_{a}'\left(a\right),
\]
which follows readily by verifying that
\[
\frac{\sqrt{\psi_{a}'\left(a\right)}\sqrt{\psi_{a}'\left(z\right)}}{\psi_{a}\left(z\right)}-\frac{1}{z-a}\underset{z\to a}{\longrightarrow}0.
\]
To prove this, notice that since 
\[
\frac{\sqrt{\psi_{a}'\left(a\right)}\sqrt{\psi_{a}'\left(z\right)}}{\psi_{a}\left(z\right)}=\frac{1}{z-a}+A+O\left(z-a\right)\,\,\,\, z\to a,
\]
by squaring this expression, it suffices to check that the residue
of $\psi_{a}'\left(z\right)/\psi_{a}^{2}\left(z\right)$ at $z=a$
vanishes. By contour integrating on a small circle $C$ around $a$
and using change of variable formula (since $\psi_{a}$ is conformal),
we have
\[
\frac{4\pi i\cdot A}{\psi_{a}'\left(a\right)}=\oint_{C}\frac{\psi_{a}'\left(z\right)dz}{\psi_{a}\left(z\right)^{2}}=\oint_{\psi_{a}^{-1}\left(C\right)}\frac{dw}{w^{2}}=0.
\]

\subsection{Proof of convergence}

The rest of this paper is devoted to the proof of the key theorem
(Theorem \ref{thm:key-thm}). This proof consists mainly of two parts:
\begin{itemize}
\item The analysis of the discrete fermionic spinors $f_{\Omega_{\delta}}$
and $f_{\mathbb{C}_{\delta}}$ as functions of their second variable
(Section 2):

\begin{itemize}
\item We prove that $f_{\Omega_{\delta}}\left(a,\cdot\right)$ and $f_{\mathbb{C}_{\delta}}\left(a,\cdot\right)$
are discrete holomorphic (in a specific sense) on $\mathcal{V}_{\Omega_{\delta}^{M}}\setminus\left\{ a\right\} $
and $\mathcal{V}_{\mathbb{C}_{\delta}^{M}}\setminus\left\{ a\right\} $
respectively (Propositions \ref{pro:bounded-domain-observable-s-holomorphicity}
and \ref{pro:full-plane-observable-s-holomorphicity}).
\item We show that $f_{\Omega_{\delta}}\left(a,\cdot\right)$ and $f_{\mathbb{C}_{\delta}}\left(a,\cdot\right)$
have the same discrete singularity at $a$: their difference $\left(f_{\Omega_{\delta}}-f_{\mathbb{C}_{\delta}}\right)\left(a,\cdot\right)$
is hence discrete holomorphic on $\mathcal{V}_{\Omega_{\delta}^{M}}$
(Propositions \ref{pro:discrete-singularity-bounded-domain-obs},
\ref{pro:full-plane-observable-singularity} and \ref{pro:boundary-effect-observable-s-holomorphicity})
\item We observe that $f_{\Omega_{\delta}}\left(a,\cdot\right)$ has some
specific boundary values on $\partial_{0}\mathcal{V}_{\Omega_{\delta}^{M}}$
(Proposition \ref{pro:boundary-condition}).
\end{itemize}
\item The proof of convergence of functions $\frac{1}{\delta}f_{\Omega_{\delta}}$,
$\frac{1}{\delta}f_{\mathbb{C}_{\delta}}$, $\frac{1}{\delta}\left(f_{\Omega_{\delta}}-f_{\mathbb{C}_{\delta}}\right)$
(Section 3):

\begin{itemize}
\item The convergence of $\frac{1}{\delta}f_{\mathbb{C}_{\delta}}$ follows
directly from the convergence result of Kenyon for the dimer coupling
function (Theorem \ref{thm:full-plane-spinor-convergence}).
\item We show that the family of functions $\left(\frac{1}{\delta}\left(f_{\Omega_{\delta}}-f_{\mathbb{C}_{\delta}}\right)\right)_{\delta>0}$
is precompact on the compact subsets of $\Omega\times\Omega$: it
admits convergent subsequential limits as $\delta\to0$ (Proposition
\ref{pro:precompactness}). Hence $\left(\frac{1}{\delta}f_{\Omega_{\delta}}\right)_{\delta>0}$
is also precompact on the compact subsets of $\Omega\times\Omega$
away from the diagonal.
\item We identify the $\delta\to0$ subsequential limits of $\frac{1}{\delta}f_{\Omega_{\delta}}$
with the function $f_{\Omega}$ (Proposition \ref{pro:identification-subsequential-limits}).
This allows to conclude the proof of Theorem \ref{thm:key-thm}.
\end{itemize}
\end{itemize}

\subsubsection*{Acknowledgements}

This research was supported by the Swiss N.S.F., the European Research
Council AG CONFRA, by EU RTN CODY, by the Chebyshev Laboratory (Department
of Mathematics and Mechanics, St.-Petersburg State University) under
RF governement grant 11.G34.31.0026 and by the National Science Foundation
under grant DMS-1106588. 

The authors would like to thank D. Chelkak for advice and useful remarks,
as well as V. Beffara, Y. Velenik, K. Kytölä, K. Izyurov, C. Boutillier,
B. de Tilière, for useful discussions. The first author would like
to thank W. Werner for his invitation to Ecole normale supérieure,
during which part of this work was completed.

\section{Analysis of the Discrete Fermionic Spinors}

In this section, we study the properties of the discrete fermionic
spinors $f_{\Omega_{\delta}}$ and $f_{\mathbb{C}_{\delta}}$ that
follow from their constructions. In the next section, we will use
these properties to prove Theorem \ref{thm:key-thm}. 

We study both spinors as functions $f_{\Omega_{\delta}}\left(a,\cdot\right)$
and $f_{\mathbb{C}_{\delta}}\left(a,\cdot\right)$ of their second
variable, keeping fixed the medial vertex $a\in\mathcal{V}_{\Omega_{\delta}^{M}}$
(which is the midpoint of a horizontal edge of $\Omega_{\delta}$). 

Let us first introduce the discrete versions of the differential operators
$\overline{\partial}$ and $\Delta$ that will be useful in this paper:
for a $\mathbb{C}$-valued function $f$, we define, wherever it makes
sense (i.e. for vertices of $\mathcal{V}_{\Omega_{\delta}}\cup\mathcal{V}_{\Omega_{\delta}^{*}}$),
\begin{eqnarray*}
\overline{\partial}_{\delta}f\left(x\right) & := & f\left(x+\frac{\delta}{2}\right)-f\left(x-\frac{\delta}{2}\right)+i\left(f\left(x+\frac{i\delta}{2}\right)-f\left(x-\frac{i\delta}{2}\right)\right),\\
\Delta_{\delta}f\left(x\right) & := & f\left(x+\delta\right)+f\left(x+i\delta\right)+f\left(x-\delta\right)+f\left(x-i\delta\right)-4f\left(x\right).
\end{eqnarray*}
In the case where one has that a vertex $y\in\left\{ x\pm\delta,x\pm i\delta\right\} $
belongs to $\partial\mathcal{V}_{\Omega_{\delta}}$ in the definition
of $\Delta_{\delta}$, the boundary vertex is the one identified with
the edge $\left\langle x,y\right\rangle \in\partial\mathcal{E}_{\Omega_{\delta}}$. 

If $e=\vec{xy}\in\vec{\mathcal{E}}_{\Omega_{\delta}}$ is an oriented
edge with $x,y\in\overline{\mathcal{V}}_{\Omega_{\delta}}$ and $f$
is a function $\overline{\mathcal{V}}_{\Omega_{\delta}}\to\mathbb{C}$
we denote by $\partial_{e}f$ the discrete partial derivative defined
by $\partial_{e}f:=f\left(y\right)-f\left(x\right)$.

\subsection{Discrete holomorphicity}

It turns out that the functions $f_{\Omega_{\delta}}\left(a,\cdot\right)$
and $f_{\mathbb{C}_{\delta}}\left(a,\cdot\right)$ are discrete holomorphic
in a specific sense, which we call \emph{s-holomorphicity} or \emph{spin-holomorphicity}. 

Let us first define this notion. With any medial edge $e\in\mathcal{E}_{\mathbb{C}_{\delta}^{M}}$,
we associate a line $\ell\left(e\right)\subset\mathbb{C}$ of the
complex plane defined by
\[
\ell\left(e\right):=\left(d-v\right)^{-\frac{1}{2}}\mathbb{R}=\left\{ \left(d-v\right)^{-\frac{1}{2}}t:t\in\mathbb{R}\right\} ,
\]
where $v\in\mathcal{V}_{\mathbb{C}_{\delta}}$ is the closest vertex
to $e$ and $d\in\mathcal{V}_{\mathbb{C}_{\delta}^{*}}$ is the closest
dual vertex to $e$. On the square lattice, the four possible lines
that we obtain are $e^{\pm\frac{\pi i}{8}}\mathbb{R}$ and $e^{\pm\frac{3\pi i}{8}}\mathbb{R}$.
When $\ell:=e^{i\theta}\mathbb{R}$ is a line in the complex plane
passing through the origin, let us denote by $\mathsf{P}_{\ell}$
the orthogonal projection on $\ell$, defined by 
\[
\mathsf{P}_{\ell}\left[z\right]:=\frac{1}{2}\left(z+e^{2i\theta}\overline{z}\right)\,\,\,\,\forall z\in\mathbb{C}.
\]

\begin{figure}
\includegraphics[width=4cm]{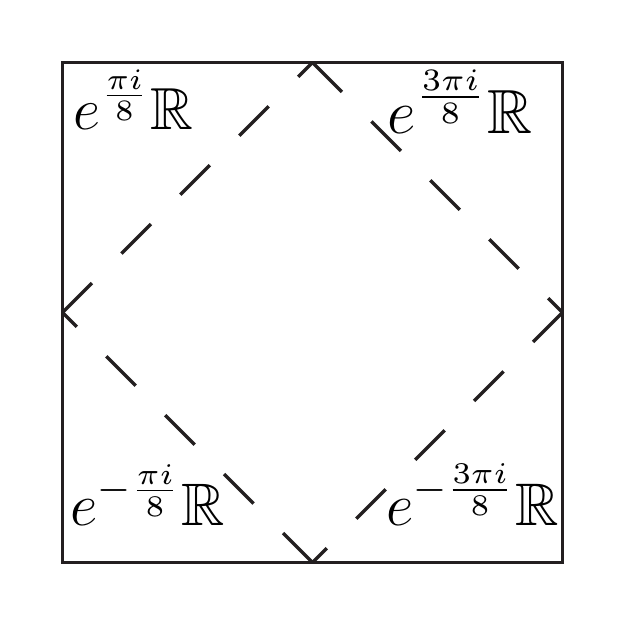}

\caption{The lines associated with the medial edges of $\Omega_{\delta}$.}
\end{figure}

\begin{defn}
\label{def:s-hol}Let $\mathcal{M}_{\delta}\subset\mathcal{V}_{\mathbb{C}_{\delta}^{M}}$
be a collection of medial vertices. We say that $f:\mathcal{M}_{\delta}\to\mathbb{C}$
is \emph{s-holomorphic on $\mathcal{M}_{\delta}$} if for any two
medial vertices $x,y\in\mathcal{M}_{\delta}$ that are adjacent in
$\mathbb{C}_{\delta}^{M}$,
\[
\mathsf{P}_{\ell\left(e\right)}\left[f\left(x\right)\right]=\mathsf{P}_{\ell\left(e\right)}\left[f\left(y\right)\right],
\]
where $e=\left\langle x,y\right\rangle \in\mathcal{E}_{\mathbb{C}_{\delta}^{M}}$. \end{defn}
\begin{rem}
Our definition is the same as the one introduced in \cite{smirnov-ii},
except that the lines that we consider are rotated by a phase of $e^{\frac{\pi i}{8}}$,
and that our lattice is rotated by an angle of $\frac{\pi}{4}$, compared
to the definitions there. In \cite{chelkak-smirnov-ii} this definition
is also used, in the more general context of isoradial graphs. 
\end{rem}

\begin{rem}
The definition of s-holomorphicity implies that a discrete version
of the Cauchy-Riemann equations is satisfied: if $f:\mathcal{M}_{\delta}\subset\mathcal{V}_{\mathbb{C}_{\delta}^{M}}\to\mathbb{C}$
is s-holomorphic and $v\in\mathcal{V}_{\mathbb{C}_{\delta}}\cup\mathcal{V}_{\mathbb{C}_{\delta}^{*}}$
is such that the four medial vertices $v\pm\frac{\delta}{2},v\pm\frac{i\delta}{2}$
are in $\mathcal{M}_{\delta}$, then we have: 
\[
\overline{\partial}_{\delta}f\left(v\right)=0.
\]
This can be found in \cite{smirnov-ii} (it follows by taking a linear
combination of the four s-holomorphicity relations between the values
$f\left(v\pm\frac{\delta}{2}\right),f\left(v\pm\frac{i\delta}{2}\right)$),
as well as the fact that satisfying this difference equation is strictly
weaker than being s-holomorphic. 
\end{rem}

\begin{figure}
\includegraphics[clip,scale=0.4]{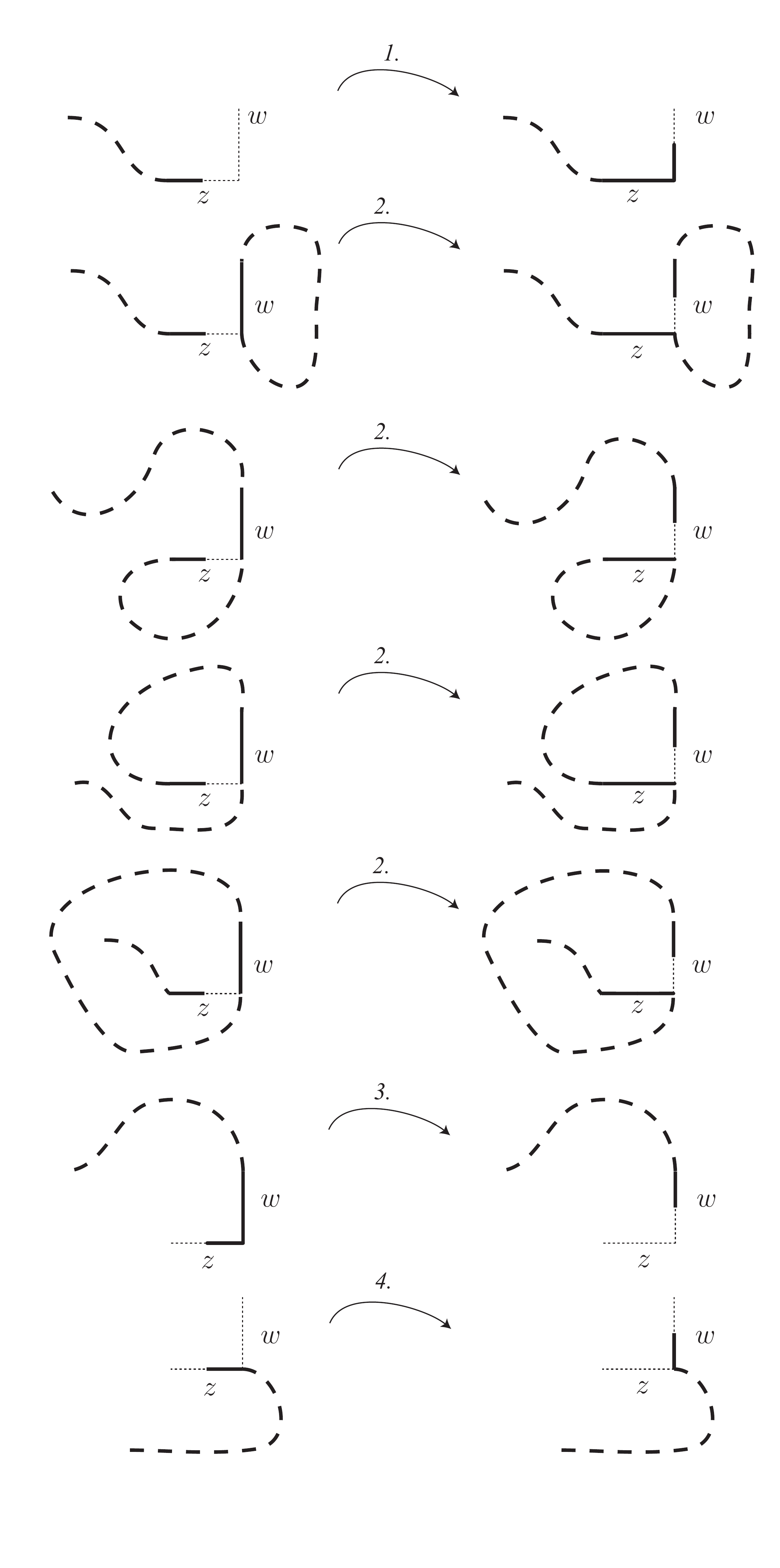}\caption{\label{fig:bijections}The four possible cases in the proof of Proposition
\ref{pro:bounded-domain-observable-s-holomorphicity}: how a configuration
is changed after two half-edges between $z$ and $w$ are removed
or added.}
\end{figure}

\begin{rem}
\label{rem:discrete-integral}If $\lambda_{\delta}=\left\{ \overrightarrow{v_{i}v_{i+1}}\in\vec{\mathcal{E}}_{\Omega_{\delta}^{M}}:i\in\mathbb{Z}/n\mathbb{Z}\right\} $
is a simple counterclockwise-oriented closed discrete contour of medial
edges and $\Lambda_{\delta}$ is the collection of points in $\mathcal{V}_{\Omega_{\delta}}\cup\mathcal{V}_{\Omega_{\delta}^{*}}$
surrounded by $\lambda_{\delta}$, then it is easy to check that for
any function $f:\mathcal{V}_{\Omega_{\delta}^{M}}\to\mathbb{C}$,
we have 
\[
\sum_{\overrightarrow{v_{i}v_{i+1}}\in\lambda_{\delta}}\frac{f\left(v_{i}\right)+f\left(v_{i+1}\right)}{2}\left(v_{i+1}-v_{i}\right)=i\delta\sum_{z\in\Lambda_{\delta}}\overline{\partial}_{\delta}f\left(v\right).
\]
In particular this sum vanishes if $f$ is discrete holomorphic.\end{rem}
\begin{prop}
\label{pro:bounded-domain-observable-s-holomorphicity}The function
$f_{\Omega_{\delta}}\left(a,\cdot\right)$ is s-holomorphic on $\mathcal{V}_{\Omega_{\delta}^{M}}\setminus\left\{ a\right\} $. \end{prop}
\begin{proof}
Let $z,w\in\mathcal{V}_{\Omega_{\delta}^{M}}\setminus\left\{ a\right\} $
be two adjacent medial vertices and let $e\in\mathcal{E}_{\Omega_{\delta}^{M}}$
be the medial edge linking them. Suppose that $z$ is the midpoint
of a horizontal edge and that $w$ is the midpoint of a vertical edge.
We prove the result in the case where $w=z+\frac{1+i}{2}\delta$ (the
other ones are symmetric). Denote by $h$ the half-edge between $z$
and $z+\frac{\delta}{2}\in\mathcal{V}_{\Omega_{\delta}}$ and by $\tilde{h}$
the half-edge between $z+\frac{\delta}{2}$ and $w$. For any $\gamma\in\mathcal{C}_{\Omega_{\delta}}\left(a,z\right)$,
define $\varphi\left(\tilde{\gamma}\right):=\gamma\oplus h\oplus\tilde{h}$
as the symmetric difference of $\gamma$ with $\left\{ h,\tilde{h}\right\} $:
if $h$ is not in $\gamma$, add it, otherwise remove it, and similarly
for $\tilde{h}$. Clearly, $\varphi$ is an involution mapping $\mathcal{C}_{\Omega_{\delta}}\left(a,z\right)$
to $\mathcal{C}_{\Omega_{\delta}}\left(a,w\right)$ and vice versa.
Moreover, for $\gamma\in\mathcal{C}_{\Omega_{\delta}}\left(a,z\right)$,
we have 
\begin{equation}
\mathsf{P}_{e^{-\frac{3\pi i}{8}}\mathbb{R}}\left[\alpha^{\left|\gamma\right|}e^{-\frac{i}{2}\mathbf{W}\left(\gamma\right)}\right]=\mathsf{P}_{e^{-\frac{3\pi i}{8}}\mathbb{R}}\left[\alpha^{\left|\varphi\left(\gamma\right)\right|}e^{-\frac{i}{2}\mathbf{W}\left(\varphi\left(\gamma\right)\right)}\right].\label{eq:bijection-proj}
\end{equation}
This identity follows from considering the four possible cases, as
shown in Figure \ref{fig:bijections}:
\begin{enumerate}
\item If $h\notin\gamma$ and $\tilde{h}\notin\gamma$, then we have $e^{-\frac{i}{2}\mathbf{W}\left(\gamma\right)}\in\mathbb{R}$,
$\left|\varphi\left(\gamma\right)\right|=\left|\gamma\right|+1$ and
$e^{-\frac{i}{2}\mathbf{W}\left(\varphi\left(\gamma\right)\right)}=e^{-\frac{\pi i}{4}}e^{-\frac{i}{2}\mathbf{W}\left(\gamma\right)}$.
\item If $h\notin\gamma$ and $\tilde{h}\in\gamma$, we have $e^{-\frac{i}{2}\mathbf{W}\left(\gamma\right)}\in\mathbb{R}$,
$\left|\varphi\left(\gamma\right)\right|=\left|\gamma\right|$ and
$e^{-\frac{i}{2}\mathbf{W}\left(\varphi\left(\gamma\right)\right)}=e^{-\frac{3\pi i}{4}}e^{-\frac{i}{2}\mathbf{W}\left(\gamma\right)}$:
there are a number of subcases, as shown in Figure \ref{fig:bijections},
for which these relations are satisfied.
\item If $h\in\gamma$ and $\tilde{h}\in\gamma$, we have $e^{-\frac{i}{2}\mathbf{W}\left(\gamma\right)}\in i\mathbb{R}$,
$\left|\varphi\left(\gamma\right)\right|=\left|\gamma\right|-1$ and
$e^{-\frac{i}{2}\mathbf{W}\left(\varphi\left(\gamma\right)\right)}=e^{-\frac{\pi i}{4}}e^{-\frac{i}{2}\mathbf{W}\left(\gamma\right)}$:
in this case, we can always choose an admissible walk on $\gamma$
that is like in Figure \ref{fig:bijections}.
\item If $h\in\gamma$ and $\tilde{h}\notin\gamma$, we have $e^{-\frac{i}{2}\mathbf{W}\left(\gamma\right)}\in i\mathbb{R}$,
$\left|\varphi\left(\gamma\right)\right|=\left|\gamma\right|$ and
$e^{-\frac{i}{2}\mathbf{W}\left(\varphi\left(\gamma\right)\right)}=e^{\frac{\pi i}{4}}e^{-\frac{i}{2}\mathbf{W}\left(\gamma\right)}$.
\end{enumerate}
In all the four cases, it is then straightforward to check that equation
\ref{eq:bijection-proj} is satisfied. By definition of $f_{\Omega_{\delta}}$
(Section \ref{def:disc-hol-spin-bded-domain}), we finally deduce
\begin{eqnarray*}
\mathsf{P}_{e^{-\frac{3\pi i}{8}}\mathbb{R}}\left[f_{\Omega_{\delta}}\left(a,z\right)\right] & = & \frac{1}{\mathbf{Z}_{\Omega_{\delta}}}\sum_{\gamma\in\mathcal{C}_{\Omega_{\delta}}\left(a,z\right)}\mathsf{P}_{e^{-\frac{3\pi i}{8}}\mathbb{R}}\left[\alpha^{\left|\gamma\right|}e^{-\frac{i}{2}\mathbf{W}\left(\gamma\right)}\right]\\
 & = & \frac{1}{\mathbf{Z}_{\Omega_{\delta}}}\sum_{\tilde{\gamma}\in\mathcal{C}_{\Omega_{\delta}}\left(a,w\right)}\mathsf{P}_{e^{-\frac{3\pi i}{8}}\mathbb{R}}\left[\alpha^{\left|\tilde{\gamma}\right|}e^{-\frac{i}{2}\mathbf{W}\left(\tilde{\gamma}\right)}\right]\\
 & = & \mathsf{P}_{e^{-\frac{3\pi i}{8}}\mathbb{R}}\left[f_{\Omega_{\delta}}\left(a,w\right)\right],
\end{eqnarray*}
which is the s-holomorphicity equation.
\end{proof}

The full-plane discrete spinor is also s-holomorphic:
\begin{prop}
\label{pro:full-plane-observable-s-holomorphicity}The function $f_{\mathbb{C}_{\delta}}\left(a,\cdot\right)$
is s-holomorphic on $\mathcal{V}_{\mathbb{C}_{\delta}^{M}}\setminus\left\{ a\right\} $.\end{prop}
\begin{proof}
This follows directly from the definition of $f_{\mathbb{C}_{\delta}}$
(Definition \ref{def:full-plane-spinor}) and of Lemma \ref{lem:full-plane-obs-s-hol}
in Appendix B.
\end{proof}

\subsection{Singularity}

Near the medial vertex $a$, the functions $f_{\Omega_{\delta}}\left(a,\cdot\right)$
and $f_{\mathbb{C}_{\delta}}\left(a,\cdot\right)$ are not s-holomorphic:
they both have a discrete singularity, but of the same nature, and
consequently the difference $\left(f_{\Omega_{\delta}}-f_{\mathbb{C}_{\delta}}\right)\left(a,\cdot\right)$
is s-holomorphic on $\mathcal{V}_{\Omega_{\delta}^{m}}$, including
the point $a$. Rather than defining the notion of discrete singularity,
let us simply describe the relations that these functions satisfy
near $a$. For $x\in\left\{ \pm1\pm i\right\} $, let us denote by
$a_{x}:=a+\frac{x\delta}{2}\in\mathcal{V}_{\Omega_{\delta}^{M}}$
the medial vertex adjacent to $a$, by $e_{x}\in\mathcal{E}_{\Omega_{\delta}^{M}}$
the medial edge between $a$ and $a_{x}$ and by $\ell_{x}$ the line
$\ell\left(e_{x}\right)$ (as in Definition \ref{def:s-hol}). Let
$e\in\mathcal{E}_{\Omega_{\delta}}^{h}$ be the horizontal edge with
midpoint $a$. Recall that $f_{\Omega_{\delta}}\left(a,a\right)$
is defined as $\mathbf{Z}_{\Omega_{\delta}}^{\left\{ e+\right\} }/\mathbf{Z}_{\Omega_{\delta}}$
(see Definition \ref{def:disc-hol-spin-bded-domain}).
\begin{prop}
\label{pro:discrete-singularity-bounded-domain-obs}Near $a$, the
function $f_{\Omega_{\delta}}\left(a,\cdot\right)$ satisfies the
relations
\begin{eqnarray*}
\mathsf{P}_{\ell_{1+i}}\left[f_{\Omega_{\delta}}\left(a,a\right)\right] & = & \mathsf{P}_{\ell_{1+i}}\left[f_{\Omega_{\delta}}\left(a,a_{1+i}\right)\right],\\
\mathsf{P}_{\ell_{1-i}}\left[f_{\Omega_{\delta}}\left(a,a\right)\right] & = & \mathsf{P}_{\ell_{1-i}}\left[f_{\Omega_{\delta}}\left(a,a_{1-i}\right)\right],\\
\mathsf{P}_{\ell_{-1+i}}\left[f_{\Omega_{\delta}}\left(a,a\right)-1\right] & = & \mathsf{P}_{\ell_{-1+i}}\left[f_{\Omega_{\delta}}\left(a,a_{-1+i}\right)\right],\\
\mathsf{P}_{\ell_{-1-i}}\left[f_{\Omega_{\delta}}\left(a,a\right)-1\right] & = & \mathsf{P}_{\ell_{-1-i}}\left[f_{\Omega_{\delta}}\left(a,a_{-1-i}\right)\right].
\end{eqnarray*}
\end{prop}
\begin{proof}
The first two relations are the s-holomorphicity relations and they
are obtained in exactly the same way as the s-holomorphicity relations
away from $a$. Indeed, let us take the same notation as in the proof
of Proposition \ref{pro:bounded-domain-observable-s-holomorphicity}
and consider the involutions $\varphi_{1+i}:\mathcal{C}_{\Omega_{\delta}}^{\left\{ e+\right\} }\to\mathcal{C}_{\Omega_{\delta}}\left(a,a_{1+i}\right)$,
and $\varphi_{1-i}:\mathcal{C}_{\Omega_{\delta}}^{\left\{ e+\right\} }\to\mathcal{C}_{\Omega_{\delta}}\left(a,a_{1-i}\right)$
defined by $\varphi_{1+i}\left(\gamma\right):=\gamma\oplus\left\langle a,a_{1}\right\rangle \oplus\left\langle a_{1},a_{1+i}\right\rangle $
and $\varphi_{1-i}\left(\gamma\right):=\gamma\oplus\left\langle a,a_{1}\right\rangle \oplus\left\langle a_{1},a_{1-i}\right\rangle $
respectively: as in the proof of Proposition \ref{pro:discrete-singularity-bounded-domain-obs},
we have that these involutions preserve the projections on $\ell_{1+i}$
and $\ell_{1-i}$ respectively (a configuration $\gamma\in\mathcal{C}_{\Omega_{\delta}}^{\left\{ e+\right\} }$
is interpreted as a configuration of winding $0$). 

For the last two relations, we have that the involutions $\varphi_{-1+i}:\mathcal{C}_{\Omega_{\delta}}^{\left\{ e-\right\} }\to\mathcal{C}_{\Omega_{\delta}}\left(a,a_{-1+i}\right)$
and $\varphi_{-1-i}:\mathcal{C}_{\Omega_{\delta}}^{\left\{ e-\right\} }\to\mathcal{C}_{\Omega_{\delta}}\left(a,a_{-1-i}\right)$,
respectively defined by $\varphi_{-1+i}\left(\gamma\right):=\gamma\oplus\left\langle a,a_{-1}\right\rangle \oplus\left\langle a_{-1},a_{-1+i}\right\rangle $
and $\varphi_{-1-i}\left(\gamma\right):=\gamma\oplus\left\langle a,a_{-1}\right\rangle \oplus\left\langle a_{-1},a_{-1-i}\right\rangle $,
are such that for any $\gamma\in\mathcal{C}_{\Omega_{\delta}}^{\left\{ e-\right\} }$,
we have
\begin{eqnarray*}
-\mathsf{P}_{\ell_{-1+i}}\left[\alpha^{\left|\gamma\right|}\right]=\mathsf{P}_{\ell_{-1+i}}\left[\alpha^{\left|\gamma\right|}e^{-\frac{i\mathbf{W}\left(\gamma\right)}{2}}\right] & = & \mathsf{P}_{\ell_{-1+i}}\left[\alpha^{\left|\varphi_{-1+i}\left(\gamma\right)\right|}e^{-\frac{i\mathbf{W}\left(\varphi_{-1+i}\left(\gamma\right)\right)}{2}}\right],\\
-\mathsf{P}_{\ell_{-1-i}}\left[\alpha^{\left|\gamma\right|}\right]=\mathsf{P}_{\ell_{-1-i}}\left[\alpha^{\left|\gamma\right|}e^{-\frac{i\mathbf{W}\left(\gamma\right)}{2}}\right] & = & \mathsf{P}_{\ell_{-1-i}}\left[\alpha^{\left|\varphi_{-1-i}\left(\gamma\right)\right|}e^{-\frac{i\mathbf{W}\left(\varphi_{-1-i}\left(\gamma\right)\right)}{2}}\right],
\end{eqnarray*}
where $\gamma$ is interpreted as a configuration with a path from
$a$ to $a$ that makes a loop, of winding number $\pm2\pi$. This
follows from the same considerations as in the proof of Proposition
\ref{pro:bounded-domain-observable-s-holomorphicity}. Hence, since
$\mathbf{Z}_{\Omega_{\delta}}^{\left\{ e-\right\} }/\mathbf{Z}_{\Omega_{\delta}}=1-f_{\Omega_{\delta}}\left(a,a\right)$,
we obtain, by summing the above equations over all $\gamma\in\mathcal{C}_{\Omega_{\delta}}^{\left\{ e-\right\} }$,
the last two identities of Proposition \ref{pro:discrete-singularity-bounded-domain-obs}.
\end{proof}
The function $f_{\mathbb{C}_{\delta}}\left(a,\cdot\right)$ has the
same type of discrete singularity as $f_{\Omega_{\delta}}\left(a,\cdot\right)$:
\begin{prop}
\label{pro:full-plane-observable-singularity}Near $a$, the function
$f_{\mathbb{C}_{\delta}}\left(a,\cdot\right)$ satisfies exactly the
same projection relations as the ones satisfied by the function $f_{\Omega_{\delta}}\left(a,\cdot\right)$,
given by Proposition \ref{pro:discrete-singularity-bounded-domain-obs}.\end{prop}
\begin{proof}
See Appendix B, Proposition \ref{prop:disc-singularity-full-plane-obs}.
\end{proof}
From Propositions \ref{pro:bounded-domain-observable-s-holomorphicity},
\ref{pro:full-plane-observable-s-holomorphicity}, \ref{pro:discrete-singularity-bounded-domain-obs}
and \ref{pro:full-plane-observable-singularity}, we readily deduce
the following:
\begin{prop}
\label{pro:boundary-effect-observable-s-holomorphicity}The function
$\left(f_{\Omega_{\delta}}-f_{\mathbb{C}_{\delta}}\right)\left(a,\cdot\right):\mathcal{V}_{\Omega_{\delta}^{M}}\to\mathbb{C}$
is s-holomorphic on $\mathcal{V}_{\Omega_{\delta}^{M}}$. 
\end{prop}

\subsection{Boundary values}

\begin{figure}
\includegraphics[scale=0.3]{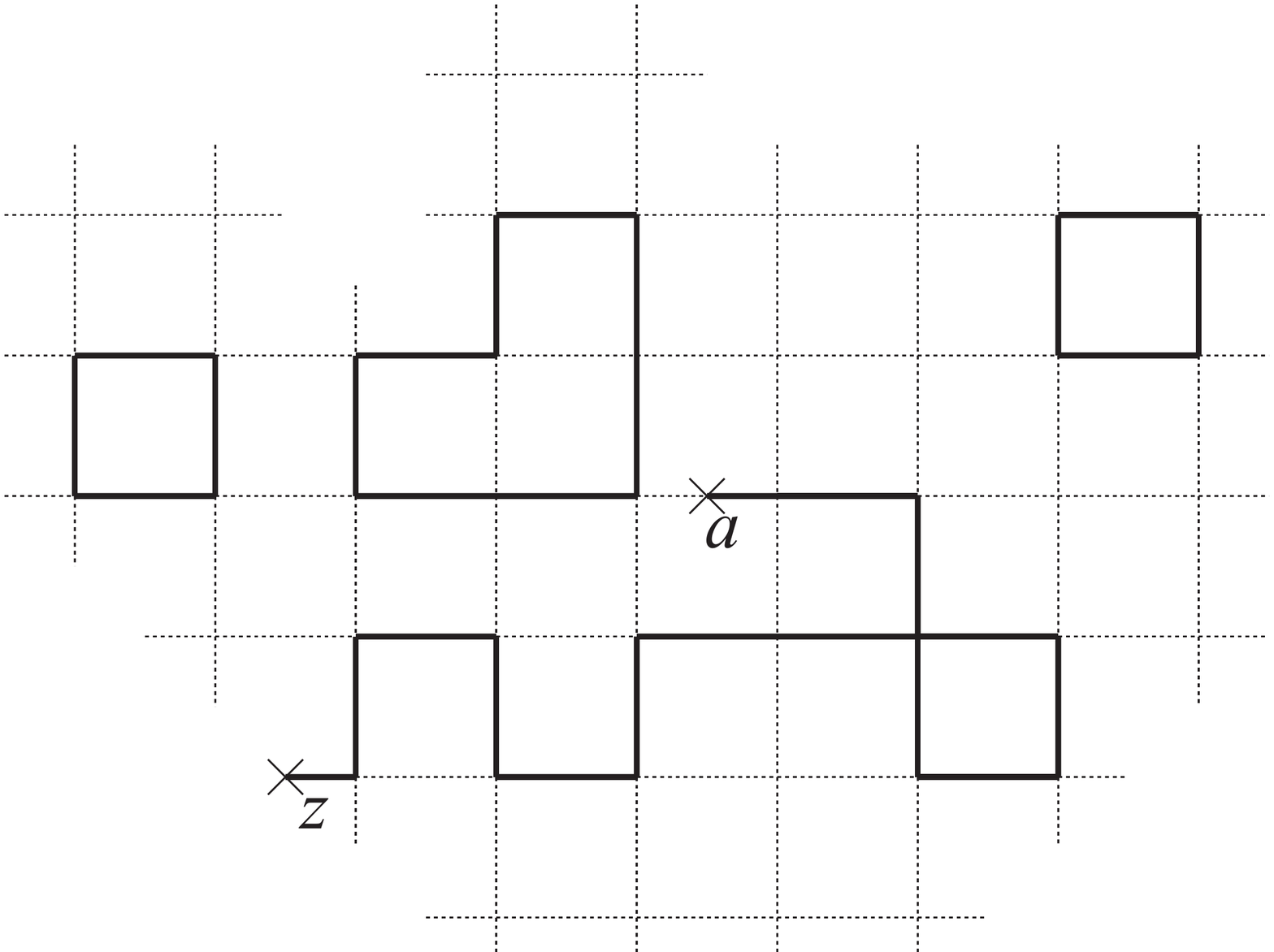}\caption{\label{fig:boundary-phase}When $z\in\partial_{0}\mathcal{V}_{\Omega_{\delta}^{M}}$,
the winding number of the walks on $\mathcal{C}_{\Omega_{\delta}}\left(a,z\right)$
is determined modulo $2\pi$. }
\end{figure}

A crucial piece of information to understand the effect of the geometry
of the discrete domain $\Omega_{\delta}$ on the average energy density
at $a\in\mathcal{V}_{\Omega_{\delta}^{M}}$ is the boundary behavior
of $f_{\Omega_{\delta}}\left(a,\cdot\right)$. On the set of boundary
medial vertices $\partial_{0}\mathcal{V}_{\Omega_{\delta}^{M}}$,
which link a vertex of $\Omega_{\delta}$ and a vertex of $\partial\Omega_{\delta}$,
the argument of $f_{\Omega_{\delta}}\left(a,\cdot\right)$ is determined
modulo $\pi$. For each $z\in\partial_{0}\mathcal{V}_{\Omega_{\delta}^{M}}$,
with $z$ being the midpoint of an edge $e\in\partial\mathcal{E}_{\Omega_{\delta}}$
between a vertex $x\in\mathcal{V}_{\Omega_{\delta}}$ and a vertex
$y\in\partial\mathcal{V}_{\Omega_{\delta}}$, denote by $\nu_{\mathrm{out}}\left(z\right)\in\partial\vec{\mathcal{E}}_{\Omega_{\delta}}$
the oriented outward-pointing edge at $z$, identified with the number
$y-x$: it is a discrete analogue of the outward-pointing normal to
the domain.
\begin{prop}
\label{pro:boundary-condition}On $\partial_{0}\mathcal{V}_{\Omega_{\delta}^{M}}$,
the argument of the value of $f_{\Omega_{\delta}}\left(a,\cdot\right)$
is determined (modulo $\pi$): for each $z\in\partial_{0}\mathcal{V}_{\Omega_{\delta}^{M}}$,
we have
\[
\Im\mathfrak{m}\left(f_{\Omega_{\delta}}\left(a,z\right)\nu_{\mathrm{out}}^{\frac{1}{2}}\left(z\right)\right)=0.
\]
\end{prop}
\begin{proof}
From topological considerations, we have that if $z\in\partial_{0}\mathcal{V}_{\Omega_{\delta}^{M}}$
and $\gamma\in\mathcal{C}_{\Omega_{\delta}}\left(a,z\right)$, then
$\Im\mathfrak{m}\left(e^{-\frac{i}{2}\mathbf{W}\left(\gamma\right)}\nu_{\mathrm{out}}^{\frac{1}{2}}\left(z\right)\right)=0$:
the winding number of any admissible walk from $a$ to $z$ is determined
modulo $2\pi$ (see Figure \ref{fig:boundary-phase}) and it is easy
to check that $e^{-\frac{i}{2}\mathbf{W}\left(\gamma\right)}$ is
a real multiple of $\left(\nu_{\mathrm{out}}\left(z\right)\right)^{-\frac{1}{2}}$.
Hence, the result follows from the definition of $f_{\Omega_{\delta}}$. 
\end{proof}

\begin{rem}
This is the same kind of Riemann-type boundary conditions as in \cite{smirnov-ii,chelkak-smirnov-ii}.
Notice that in these papers, the argument of the function on the boundary
is fully determined (not only modulo $\pi$).
\end{rem}

\subsection{Discrete integration}

An essential tool that we will use for deriving the convergence of
$f_{\Omega_{\delta}}\left(a,\cdot\right)$ is the possibility to define
a discrete version of the antiderivative of the square of an s-holomorphic
function, cf. \cite{smirnov-ii}.
\begin{prop}
\label{pro:integral-of-the-square}Let $f:\mathcal{V}_{\mathcal{D}_{\delta}^{M}}\to\mathbb{C}$
be an s-holomorphic function on a discrete domain $\mathcal{D}_{\delta}$
and let $x\in\overline{\mathcal{V}}_{\mathcal{D}_{\delta}}\cup\overline{\mathcal{V}}_{\mathcal{D}_{\delta}^{*}}$
(where $\overline{\mathcal{V}}_{\mathcal{D}_{\delta}}=\mathcal{V}_{\mathcal{D}_{\delta}}\cup\partial\mathcal{V}_{\mathcal{D}_{\delta}}$
and $\overline{\mathcal{V}}_{\mathcal{D}_{\delta}^{*}}=\mathcal{V}_{\mathcal{D}_{\delta}^{*}}\cup\partial\mathcal{V}_{\mathcal{D}_{\delta}^{*}}$).
Then there exists a (possibly multivalued) discrete analogue $\mathbb{I}_{x,\delta}\left[f\right]:\overline{\mathcal{V}}_{\mathcal{D}_{\delta}}\cup\overline{\mathcal{V}}_{\mathcal{D}_{\delta}^{*}}\to\mathbb{R}$
of the antiderivative
\[
-\Re\mathfrak{e}\left(\int_{x}f^{2}\right),
\]
uniquely defined by $\mathbb{I}_{x,\delta}\left[f\right]\left(x\right):=0$
and
\begin{eqnarray*}
\mathbb{I}_{x,\delta}\left[f\right]\left(b\right)-\mathbb{I}_{x,\delta}\left[f\right]\left(w\right) & = & \sqrt{2}\delta\left|\mathsf{P}_{\ell\left(e^{*}\right)}\left[f\left(y\right)\right]\right|^{2}=\sqrt{2}\delta\left|\mathsf{P}_{\ell\left(e^{*}\right)}\left[f\left(z\right)\right]\right|^{2}\\
 &  & \,\,\,\,\forall b\in\mathcal{V}_{\Omega_{\delta}},w\in\mathcal{V}_{\Omega_{\delta}^{*}}\,:\,\left|b-w\right|=\frac{\delta}{\sqrt{2}},
\end{eqnarray*}
where $e^{*}=\left\langle y,z\right\rangle \in\mathcal{E}_{\mathcal{D}_{\delta}^{M}}$
is the medial edge which is between $b$ and $w$. If $\mathcal{D}_{\delta}$
is simply connected, then the function $\mathbb{I}_{x,\delta}\left[f\right]$
is globally well-defined (single-valued). When the choice of the point
$x$ is irrelevant, we will ommit it and simply write $\mathbb{I}_{\delta}\left[f\right]$.\end{prop}
\begin{rem}
\label{rem:integral-square}It follows from the definition of $\mathbb{I}_{\delta}\left[f\right]$
that for any pair of adjacent vertices $x,y\in\overline{\mathcal{V}}_{\mathcal{D}_{\delta}}$,
we have
\[
\mathbb{I}_{\delta}\left[f\right]\left(x\right)-\mathbb{I}_{\delta}\left[f\right]\left(y\right)=-\Re\mathfrak{e}\left(f\left(\frac{x+y}{2}\right)^{2}\left(y-x\right)\right),
\]
and similarly if $x,y\in\overline{\mathcal{V}}_{\mathcal{D}_{\delta}^{*}}$
are adjacent dual vertices. From there it is easy to see that if the
mesh size is small, $\mathbb{I}_{\delta}\left[f\right]$ is a good
approximation of $-\Re\mathfrak{e}\left(\int f^{2}\right)$.
\end{rem}
We denote by $\mathbb{I}_{x,\delta}^{\bullet}\left[f\right]$ and
$\mathbb{I}_{x,\delta}^{\circ}\left[f\right]$ the restrictions of
$\mathbb{I}_{x,\delta}\left[f\right]$ to $\overline{\mathcal{V}}_{\mathcal{D}_{\delta}}$
and $\overline{\mathcal{V}}_{\mathcal{D}_{\delta}^{*}}$ respectively.
We have the following:
\begin{prop}
\label{pro:integral-of-square-sub-super-boundary-values}The function
$\mathbb{I}_{\delta}^{\bullet}\left[f\right]:\overline{\mathcal{V}}_{\mathcal{D}_{\delta}}\to\mathbb{R}$
is discrete subharmonic and the function $\mathbb{I}_{\delta}^{\circ}\left[f\right]:\overline{\mathcal{V}}_{\mathcal{D}_{\delta}^{*}}\to\mathbb{R}$
is discrete superharmonic: we have
\begin{eqnarray*}
\Delta_{\delta}\mathbb{I}_{\delta}^{\bullet}\left[f\right]\left(v\right) & \geq & 0\,\,\,\,\forall v\in\mathcal{V}_{\mathcal{D}_{\delta}},\\
\Delta_{\delta}\mathbb{I}_{\delta}^{\circ}\left[f\right]\left(v\right) & \leq & 0\,\,\,\,\forall v\in\mathcal{V}_{\mathcal{D}_{\delta}^{*}}.
\end{eqnarray*}
If $m\in\partial\mathcal{V}_{\mathcal{D}_{\delta}^{M}}$, then we
have 
\[
\partial_{\nu_{\mathrm{out}}\left(m\right)}\mathbb{I}_{\delta}^{\bullet}\left[f\right]=\Im\mathfrak{m}\left(f\left(m\right)\nu_{\mathrm{out}}^{\frac{1}{2}}\left(m\right)\right)^{2}-\Re\mathfrak{e}\left(f\left(m\right)\nu_{\mathrm{out}}^{\frac{1}{2}}\left(m\right)\right)^{2},
\]
where $\nu_{\mathrm{out}}\left(m\right)\in\partial\vec{\mathcal{E}}_{\mathcal{D}_{\delta}}$
is the oriented edge from $a\in\mathcal{D}_{\delta}$ to $b\in\partial\mathcal{D}_{\delta}$,
the midpoint of which is $m$. \end{prop}
\begin{proof}
For the subharmonicity/superharmonicity deduced from the s-holomorphicity
of $f$, see Lemma 3.8 in \cite{smirnov-ii} (the fact that the phases
are different does not affect the result). The normal derivative statement
follows directly from the definition of $\mathbb{I}_{\delta}\left[f\right]$.
\end{proof}
In the case of the discrete fermionic spinor $f_{\Omega_{\delta}}\left(\cdot,\cdot\right)$,
the boundary condition for $\mathbb{I}_{\delta}\left[f_{\Omega_{\delta}}\left(a,\cdot\right)\right]\left(\cdot\right)$
becomes particularly simple.
\begin{prop}
\label{pro:constant-value-integral-square}The function $\mathbb{I}_{\delta}^{\circ}\left[f_{\Omega_{\delta}}\left(a,\cdot\right)\right]:\overline{\mathcal{V}}_{\Omega_{\delta}^{*}}\to\mathbb{R}$
is constant on $\partial\mathcal{V}_{\Omega_{\delta}^{*}}$ and for
each $m\in\partial_{0}\mathcal{V}_{\Omega_{\delta}^{M}}$, 
\[
\partial_{\nu_{\mathrm{out}}\left(m\right)}\mathbb{I}_{\delta}^{\bullet}\left[f_{\Omega_{\delta}}\left(a,\cdot\right)\right]=-\left|f_{\Omega_{\delta}}\left(a,m\right)\right|^{2}.
\]
\end{prop}
\begin{proof}
The first statement follows from the construction of $\mathbb{I}_{\delta}^{\circ}\left[f_{\Omega_{\delta}}\left(a,\cdot\right)\right]$
and from the boundary condition for $f_{\Omega_{\delta}}$ (Proposition
\ref{pro:boundary-condition}).

The statement for $\mathbb{I}_{\delta}^{\bullet}\left[f_{\Omega_{\delta}}\left(a,\cdot\right)\right]$
follows directly from Proposition \ref{pro:integral-of-square-sub-super-boundary-values}
and the boundary condition for $f_{\Omega_{\delta}}$ (Proposition
\ref{pro:boundary-condition} again).\end{proof}
\begin{rem}
Note that $\mathbb{I}_{\delta}\left[f_{\Omega_{\delta}}\left(a,\cdot\right)\right]$
is single-valued (as follows from Proposition \ref{pro:constant-value-integral-square})
and well-defined on $\overline{\mathcal{V}}_{\Omega_{\delta}}\cup\overline{\mathcal{V}}_{\Omega_{\delta}^{*}}$
but that the presence of a singularity near $a$ implies that and
$\mathbb{I}_{\delta}^{\bullet}\left[f_{\Omega_{\delta}}\left(a,\cdot\right)\right]$
and $\mathbb{I}_{\delta}^{\circ}\left[f_{\Omega_{\delta}}\left(a,\cdot\right)\right]$
are (at least a priori) not subharmonic or superharmonic near $a$
(more precisely at $a\pm\frac{\delta}{2},a\pm\frac{i\delta}{2}$). 
\end{rem}

\section{Convergence of the Discrete Fermionic Spinors}

We now turn to the convergence of the three functions $\frac{1}{\delta}f_{\Omega_{\delta}},\frac{1}{\delta}f_{\mathbb{C}_{\delta}},\frac{1}{\delta}\left(f_{\Omega_{\delta}}-f_{\mathbb{C}_{\delta}}\right)$
as $\delta\to0$ (Theorem \ref{thm:key-thm}). For this, we use the
discrete results derived in the previous section: the s-holomorphicity,
the discrete singularity and the boundary values. As we will discuss
convergence questions, we will always, when necessary, identify the
points of the complex plane with the closest vertices on the graphs
considered. In this way, we will extend functions defined on the vertices
of the graphs $\Omega_{\delta},\Omega_{\delta}^{*},\Omega_{\delta}^{m}$
to functions defined on $\Omega$. In particular, for the discrete
holomorphic spinors, when we write $f_{\Omega_{\delta}}\left(a,z\right)$
or $f_{\mathbb{C}_{\delta}}\left(a,z\right)$ for $a,z\in\Omega$,
we identify $a$ with the closest midpoint of a horizontal edge of
$\mathcal{E}_{\Omega_{\delta}}^{h}$ and $z$ with the closest  midpoint
of an arbitrary edge of $\mathcal{E}_{\Omega_{\delta}}$. 

The convergence of $f_{\mathbb{C}_{\delta}}$ almost immediately follows
from the work of Kenyon \cite{kenyon-i}:
\begin{thm}
\label{thm:full-plane-spinor-convergence}For any $\epsilon>0$. we
have
\[
\frac{1}{\delta}f_{\mathbb{C}_{\delta}}\left(a,z\right)\underset{\delta\to0}{\longrightarrow}f_{\mathbb{C}}\left(a,z\right),
\]
uniformly on $\left\{ \left(a,z\right)\in\mathbb{C}^{2}:\left|a-z\right|\geq\epsilon\right\} $,
where 
\[
f_{\mathbb{C}}\left(a,z\right)=\frac{1}{2\pi\left(z-a\right)}.
\]
\end{thm}
\begin{proof}
See the last paragraph of Appendix B.
\end{proof}
For the convergence of $\frac{1}{\delta}f_{\Omega_{\delta}}$ and
$\frac{1}{\delta}\left(f_{\Omega_{\delta}}-f_{\mathbb{C}_{\delta}}\right)$,
we proceed in two steps: we first show that the family of functions
$\left(\frac{1}{\delta}\left(f_{\Omega_{\delta}}-f_{\mathbb{C}_{\delta}}\right)\right)_{\delta>0}$
is precompact. Precompactness for $\left(\frac{1}{\delta}f_{\Omega_{\delta}}\right)_{\delta>0}$
will then readily follow from Theorem \ref{thm:full-plane-spinor-convergence}.
We then identify uniquely the subsequential limits of $\left(\frac{1}{\delta}f_{\Omega_{\delta}}\right)_{\delta>0}$;
this also identifies the ones of $\left(\frac{1}{\delta}\left(f_{\Omega_{\delta}}-f_{\mathbb{C}_{\delta}}\right)\right)_{\delta>0}$.

\subsection{Precompactness}

We now state our main precompactness result:
\begin{prop}
\label{pro:precompactness}The family of functions 
\[
\left(\left(a,z\right)\mapsto\frac{1}{\delta}\left(f_{\Omega_{\delta}}-f_{\mathbb{C}_{\delta}}\right)\left(a,z\right)\right)_{\delta>0}
\]
is precompact in the topology of uniform convergence on the compact
subsets of $\Omega\times\Omega$, and hence the family of functions
\[
\left(\left(a,z\right)\mapsto\frac{1}{\delta}f_{\Omega_{\delta}}\left(a,z\right)\right)_{\delta>0}
\]
is precompact in the topology of uniform convergence on the compact
subsets of $\Omega\times\Omega$ that are away from the diagonal. \end{prop}
\begin{proof}
Set $f_{\Omega_{\delta}}^{\mathbb{C}_{\delta}}:=f_{\Omega_{\delta}}-f_{\mathbb{C}_{\delta}}$.
By Proposition \ref{pro:boundary-condition}, we have that for any
$x\in\partial_{0}\mathcal{V}_{\Omega_{\delta}^{M}}$, 
\[
\Im\mathfrak{m}\left(f_{\Omega_{\delta}}^{\mathbb{C}_{\delta}}\left(a,x\right)\cdot\nu_{\mathrm{out}}^{\frac{1}{2}}\left(x\right)\right)=-\Im\mathfrak{m}\left(f_{\mathbb{C}_{\delta}}\left(a,x\right)\cdot\nu_{\mathrm{out}}^{\frac{1}{2}}\left(x\right)\right).
\]
By Theorem \ref{thm:full-plane-spinor-convergence} and the fact that
$\partial\Omega$ is smooth (and hence the number of medial vertices
in $\partial_{0}\mathcal{V}_{\Omega_{\delta}^{M}}$ is $O\left(\delta^{-1}\right))$
we deduce that the family of functions
\[
\left(a\mapsto\sum_{x\in\partial_{0}\mathcal{V}_{\Omega_{\delta}^{M}}}\Im\mathfrak{m}\left(\frac{1}{\delta}f_{\Omega_{\delta}}^{\mathbb{C}_{\delta}}\left(a,x\right)\cdot\nu_{\mathrm{out}}^{\frac{1}{2}}\left(x\right)\right)^{2}\cdot\delta\right)_{\delta>0}
\]
is uniformly bounded and equicontinuous on the compact subsets of
$\Omega$ (since $\frac{1}{\delta}f_{\mathbb{C}_{\delta}}$ is uniformly
convergent by Theorem \ref{thm:full-plane-spinor-convergence}). By
Proposition \ref{pro:uniform-control} below, we obtain that the family
of functions 
\[
\left(\frac{1}{\delta}f_{\Omega_{\delta}}^{\mathbb{C}_{\delta}}:\mathcal{V}_{\Omega_{\delta}^{M}}\times\mathcal{V}_{\Omega_{\delta}^{M}}\to\mathbb{C}\right)_{\delta>0}
\]
is uniformly bounded and equicontinuous and hence we get the desired
result by extending the functions $f_{\Omega_{\delta}}^{\mathbb{C}_{\delta}}$
in a uniformly continuous way to $\Omega\times\Omega$ (for instance
by piecewise-linear interpolation) and by using then Arzelà-Ascoli
theorem.\end{proof}
\begin{prop}
\label{pro:uniform-control}There exists a universal constant $C>0$
such that for each $\delta>0$ and any s-holomorphic function $u_{\delta}:\mathcal{V}_{\Omega_{\delta}^{M}}\to\mathbb{C}$,
we have, for any $v\in\mathcal{V}_{\Omega_{\delta}^{M}}\setminus\partial_{0}\mathcal{V}_{\Omega_{\delta}^{M}}$,
\begin{eqnarray*}
\left|u_{\delta}\left(v\right)\right| & \leq & C\sqrt{\frac{\sum_{x\in\partial_{0}\mathcal{V}_{\Omega_{\delta}^{M}}}\Im\mathfrak{m}\left(u_{\delta}\left(x\right)\cdot\nu_{\mathrm{out}}^{\frac{1}{2}}\left(x\right)\right)^{2}\cdot\delta}{\mathrm{dist}\left(v,\partial_{0}\mathcal{V}_{\Omega_{\delta}^{M}}\right)}},\\
\frac{1}{\delta}\|\nabla_{\delta}u_{\delta}\left(v\right)\|^{2} & \leq & C\sqrt{\frac{\sum_{x\in\partial_{0}\mathcal{V}_{\Omega_{\delta}^{M}}}\Im\mathfrak{m}\left(u_{\delta}\left(x\right)\cdot\nu_{\mathrm{out}}^{\frac{1}{2}}\left(x\right)\right)^{2}\cdot\delta}{\mathrm{dist}\left(v,\partial_{0}\mathcal{V}_{\Omega_{\delta}^{M}}\right)^{3}}},
\end{eqnarray*}
where $\nabla_{\delta}u_{\delta}\left(v\right)=\left(u_{\delta}\left(v+\delta\right)-u_{\delta}\left(v\right),u_{\delta}\left(v+i\delta\right)-u_{\delta}\left(v\right)\right)$. \end{prop}
\begin{proof}
Consider the function $\mathbb{I}_{\delta}\left[u_{\delta}\right]$,
normalized to be $0$ at an arbitrary point. By subharmonicity (Proposition
\ref{pro:integral-of-square-sub-super-boundary-values}), a discrete
integration by parts, and again by Proposition \ref{pro:integral-of-square-sub-super-boundary-values},
we obtain
\begin{eqnarray*}
0 & \leq & \sum_{b\in\mathcal{V}_{\Omega_{\delta}}}\Delta_{\delta}\mathbb{I}_{\delta}^{\bullet}\left[u_{\delta}\right]=\sum_{x\in\partial_{0}\mathcal{V}_{\Omega_{\delta}^{M}}}\partial_{\nu_{\mathrm{out}}\left(x\right)}\mathbb{I}_{\delta}^{\bullet}\left[u_{\delta}\right]\\
 & = & \sum_{x\in\partial_{0}\mathcal{V}_{\Omega_{\delta}^{M}}}\left(\Im\mathfrak{m}\left(u_{\delta}\left(x\right)\cdot\nu_{\mathrm{out}}^{\frac{1}{2}}\left(x\right)\right)\right)^{2}-\left(\Re\mathfrak{e}\left(u_{\delta}\left(x\right)\cdot\nu_{\mathrm{out}}^{\frac{1}{2}}\left(x\right)\right)\right)^{2}
\end{eqnarray*}
and from the last identity we also deduce that
\[
\sum_{x\in\partial_{0}\mathcal{V}_{\Omega_{\delta}^{M}}}\left|u_{\delta}\left(x\right)\right|^{2}\leq2\sum_{x\in\partial_{0}\mathcal{V}_{\Omega_{\delta}^{M}}}\left(\Im\mathfrak{m}\left(u_{\delta}\left(x\right)\cdot\nu_{\mathrm{out}}^{\frac{1}{2}}\left(x\right)\right)\right)^{2}.
\]
On the other hand, from the construction of $\mathbb{I}_{\delta}\left[u\right]$,
it is easy to see that 
\[
\max_{z\in\partial\mathcal{V}_{\Omega_{\delta}}\cup\partial\mathcal{V}_{\Omega_{\delta}^{*}}}\left|\mathbb{I}_{\delta}\left[u_{\delta}\right]\left(z\right)\right|\leq\sqrt{2}\left(\sum_{x\in\partial_{0}\mathcal{V}_{\Omega_{\delta}^{M}}}\left|u_{\delta}\left(x\right)\right|^{2}\cdot\delta\right).
\]
By subharmonicity of $\mathbb{I}_{\delta}^{\bullet}\left[u_{\delta}\right]$,
superharmonicity of $\mathbb{I}_{\delta}^{\circ}\left[u_{\delta}\right]$
and the construction of $\mathbb{I}_{\delta}\left[u_{\delta}\right]$
(Proposition \ref{pro:integral-of-the-square}), we have
\[
\max_{w\in\overline{\mathcal{V}}_{\Omega_{\delta}}\cup\overline{\mathcal{V}}_{\Omega_{\delta}^{*}}}\left|\mathbb{I}_{\delta}\left[u_{\delta}\right]\left(w\right)\right|=\max_{z\in\partial\mathcal{V}_{\Omega_{\delta}}\cup\partial\mathcal{V}_{\Omega_{\delta}^{*}}}\left|\mathbb{I}_{\delta}\left[u_{\delta}\right]\left(z\right)\right|.
\]
By Theorem 3.12 in \cite{chelkak-smirnov-ii} (the construction there
is the same as the one of our paper, up to a multiplication by an
overall complex factor, which does not affect the result), there exists
then a universal constant $\tilde{C}>0$ such that for any $v\in\mathcal{V}_{\Omega_{\delta}^{M}}\setminus\partial_{0}\mathcal{V}_{\Omega_{\delta}^{M}}$
\begin{eqnarray*}
\left|u_{\delta}\left(v\right)\right|^{2} & \leq & \tilde{C}\frac{\max_{w\in\overline{\mathcal{V}}_{\Omega_{\delta}}\cup\overline{\mathcal{V}}_{\Omega_{\delta}^{*}}}\left|\mathbb{I}_{\delta}\left[u_{\delta}\right]\left(w\right)\right|}{\mathrm{dist}\left(v,\partial_{0}\mathcal{V}_{\Omega_{\delta}^{M}}\right)},\\
\|\nabla_{\delta}u_{\delta}\left(v\right)\|^{2} & \leq & \tilde{C}\frac{\max_{w\in\overline{\mathcal{V}}_{\Omega_{\delta}}\cup\overline{\mathcal{V}}_{\Omega_{\delta}^{*}}}\left|\mathbb{I}_{\delta}\left[u_{\delta}\right]\left(w\right)\right|}{\mathrm{dist}\left(v,\partial_{0}\mathcal{V}_{\Omega_{\delta}^{M}}\right)^{3}}.
\end{eqnarray*}
We therefore deduce the desired inequalities.
\end{proof}

\subsection{Identification of the limit}

We can now uniquely identify the subsequential limits of $\left(\frac{1}{\delta}f_{\Omega_{\delta}}\right)_{\delta>0}$
as $\delta\to0$ (we will often make a slight of abuse of notation
and simply denote the family of functions by $\frac{1}{\delta}f_{\Omega_{\delta}}$).
Let us start with a characterization of the continuous fermionic spinor
$f_{\Omega}\left(a,\cdot\right)$ (defined in Section \ref{sub:convergence-results}).
\begin{lem}
\label{lem:uniqueness-continuous-observable}The function $f_{\Omega}\left(a,\cdot\right)$
is the unique holomorphic function such that 
\[
z\mapsto f_{\Omega}\left(a,z\right)-\frac{1}{2\pi\left(z-a\right)}
\]
is bounded near $z=a$ and such that 
\begin{equation}
\Im\mathfrak{m}\left(f_{\Omega}\left(a,z\right)\nu_{\mathrm{out}}^{\frac{1}{2}}\left(z\right)\right)=0\,\,\,\,\forall z\in\partial\Omega,\label{eq:bdry-cond}
\end{equation}
where $\nu_{\mathrm{out}}$ denotes the outward-pointing normal to
$\partial\Omega$. 

The boundary condition \ref{eq:bdry-cond} is equivalent to the condition
that the antiderivative $F\left(z\right)=-\Re\mathfrak{e}\left(\int^{z}f_{\Omega}^{2}\left(a,w\right)dw\right)$
is single-valued on $\Omega\setminus\left\{ a\right\} $, constant
on $\partial\Omega$ and satisfies
\[
\partial_{\nu_{\mathrm{out}}\left(z\right)}F\leq0\,\,\,\,\forall z\in\partial\Omega,
\]
where $\partial_{\nu_{\mathrm{out}}\left(z\right)}F$ denotes the
normal derivative of $F$ in the outward-pointing direction.\end{lem}
\begin{proof}
It is straightforward to check from the definition (Section \ref{sub:convergence-results})
that $f_{\Omega}\left(a,\cdot\right)$ has a simple pole of order
$1$ and residue $\frac{1}{2\pi}$ at $z=a$ and satisfies the boundary
condition \ref{eq:bdry-cond}.

Let $\tilde{f}$ be another function with the same pole and boundary
condition. Then the function $g$ defined by $g\left(z\right):=f_{\Omega}\left(a,z\right)-\tilde{f}\left(z\right)$
extends holomorphically to $\Omega$ and satisfies 
\[
\Im\mathfrak{m}\left(g\left(z\right)\nu_{\mathrm{out}}^{\frac{1}{2}}\left(z\right)\right)=0\,\,\,\,\forall z\in\partial\Omega.
\]
The function $G:\Omega\to\mathbb{C}$ defined by $G\left(w\right):=\int^{w}g^{2}\left(z\right)dz$
has constant real part on $\partial\Omega$ and hence is constant
on $\Omega$, by the maximum principle and the Cauchy-Riemann equations.
Hence $f_{\Omega}\left(a,\cdot\right)=\tilde{f}\left(\cdot\right)$. 

For the second part of the statement, notice that the boundary condition
\ref{eq:bdry-cond} implies that $f_{\Omega}^{2}\left(a,\cdot\right)\nu_{\mathrm{out}}\left(\cdot\right)$
is purely real on $\partial\Omega$, and hence $F$ must be constant
on $\partial\Omega$ (when going along the boundary, one integrates
$f_{\Omega}^{2}\left(a,\cdot\right)d\tau$, where $\tau$ is the tangent
to the boundary, which is orthogonal to the normal $\nu_{\mathrm{out}}$);
this implies that $F$ is single-valued on $\Omega\setminus\left\{ a\right\} $
(since $\Omega$ is simply connected) and that 
\[
\partial_{\mathfrak{\nu}_{\mathrm{out}}\left(z\right)}F=-\left|f_{\Omega}\left(a,z\right)\right|^{2}\,\,\,\,\forall z\in\partial\Omega.
\]
Conversely, it is easy to check that if $F$ is constant on $\partial\Omega$,
then for any $z\in\partial\Omega$, we have $f_{\Omega}^{2}\left(a,z\right)\nu_{\mathrm{out}}\left(z\right)\in\mathbb{R}$.
Moreover, for any $z\in\partial\Omega$, we have that $\partial_{\nu_{\mathrm{out}}\left(z\right)}F\leq0$
implies $f_{\Omega}^{2}\left(a,z\right)\nu_{\mathrm{out}}\left(z\right)\geq0$,
which is equivalent to $\Im\mathfrak{m}\left(f_{\Omega}\left(a,z\right)\nu_{\mathrm{out}}^{\frac{1}{2}}\left(z\right)\right)=0$. 
\end{proof}

Let us also give a lemma which will be useful to connect the discrete
spinors to the continuous ones:
\begin{lem}
\textup{\label{lem:boundedness-bdry-integral-square}We have the uniform
bound 
\[
\sup_{\delta>0}\left(\sum_{z\in\partial_{0}\mathcal{V}_{\Omega_{\delta}^{m}}}\left|f_{\Omega_{\delta}}\left(z\right)\right|^{2}\cdot\delta\right)\,<\,\infty.
\]
}\end{lem}
\begin{proof}
This follows directly from the the proof of precompactness of $\left(\frac{1}{\delta}\left(f_{\Omega_{\delta}}-f_{\mathbb{C}_{\delta}}\right)\right)_{\delta}$
(Proposition \ref{pro:precompactness}), the convergence of $\frac{1}{\delta}f_{\mathbb{C}_{\delta}}$
(Theorem \ref{thm:full-plane-spinor-convergence}) and the fact that
$\partial\Omega$ is smooth (the number of medial vertices in $\partial_{0}\mathcal{V}_{\Omega_{\delta}^{m}}$
is $O\left(\delta^{-1}\right)$).
\end{proof}

We now pass to the identification of the subsequential limits of $\frac{1}{\delta}f_{\Omega_{\delta}}\left(a,\cdot\right)$
as $\delta\to0$:
\begin{prop}
\label{pro:identification-subsequential-limits}Let $\delta_{n}$
be a sequence with $\delta_{n}\underset{n\to\infty}{\longrightarrow}0$
such that $\frac{1}{\delta_{n}}f_{\Omega_{\delta_{n}}}\left(a,\cdot\right)\underset{n\to\infty}{\longrightarrow}f\left(\cdot\right)$,
uniformly on the compact subsets of $\Omega\setminus\left\{ a\right\} $.
Then $f\left(\cdot\right)=f_{\Omega}\left(a,\cdot\right)$, where
$f_{\Omega}\left(a,\cdot\right)$ is defined in Section \ref{sub:convergence-results}.\end{prop}
\begin{proof}
For each $\delta>0$, set $f_{\delta}\left(\cdot\right):=\frac{1}{\delta}f_{\Omega_{\delta}}\left(a,\cdot\right)$. 

Let us first remark that $f\left(\cdot\right)$ is holomorphic on
$\Omega\setminus\left\{ a\right\} $, as it satisfies Morera's condition:
the integral of $f\left(\cdot\right)$ on any contractible contour
vanishes, since it can be approximated by a Riemann sum involving
$f_{\delta_{n}}\left(\cdot\right)$ (for $\delta_{n}$ small), which
vanishes identically as explained in Remark \ref{rem:discrete-integral}.
Fix a point $p\in\Omega\setminus\left\{ a\right\} $. By Lemma \ref{lem:uniqueness-continuous-observable},
to identify $f$ with $f_{\Omega}\left(a,\cdot\right)$, it suffices
to check that $F$, defined by $F\left(z\right):=-\Re\mathfrak{e}\left(\int_{p}^{z}f^{2}\left(w\right)dw\right)$,
satisfies the conditions of the second part of that lemma. 

Let $F_{\delta}:\overline{\mathcal{V}}_{\Omega_{\delta}}\cup\overline{\mathcal{V}}_{\Omega_{\delta}^{*}}\to\mathbb{R}$
be the discrete antiderivative $\mathbb{I}_{p,\delta}\left[f_{\delta}\right]$,
as defined in Proposition \ref{pro:integral-of-the-square}, and let
$F_{\delta}^{\bullet}$ and $F_{\delta}^{\circ}$ denote the restrictions
of $F_{\delta}$ to $\overline{\mathcal{V}}_{\Omega_{\delta}}$ and
$\overline{\mathcal{V}}_{\Omega_{\delta}^{*}}$, which are discrete
subharmonic and superharmonic respectively (away from $a$), by Proposition
\ref{pro:integral-of-square-sub-super-boundary-values}. By Proposition
\ref{pro:constant-value-integral-square}, the function $F_{\delta}^{\bullet}$
is constant on $\partial\mathcal{V}_{\Omega_{\delta}^{*}}$; denote
by $F_{\delta}\left(\partial\Omega\right)$ this value. Fix a smooth
doubly connected domain $\Upsilon\subset\Omega\setminus\left\{ a\right\} $
such that $\partial\Omega\subset\partial\Upsilon$ and $\mathrm{dist}\left(a,\partial\Upsilon\right)>0$
(one of the component of $\partial\Upsilon$ is $\partial\Omega$
and the other is a simple loop surrounding $a$). Let us write $F_{\delta}^{\bullet}=:H_{\delta}^{\bullet}+S_{\delta}^{\bullet}$
and $F_{\delta}^{\circ}=:H_{\delta}^{\circ}+S_{\delta}^{\circ}$,
where
\begin{itemize}
\item $H_{\delta}^{\bullet}:\overline{\mathcal{V}}_{\Upsilon_{\delta}}\to\mathbb{R}$
is discrete harmonic, with $H_{\delta}^{\bullet}:=F_{\delta}^{\bullet}$
on $\partial\mathcal{V}_{\Upsilon_{\delta}}$,
\item $S_{\delta}^{\bullet}:\overline{\mathcal{V}}_{\Upsilon_{\delta}}\to\mathbb{R}$
is discrete subharmonic, with $S_{\delta}^{\bullet}:=0$ on $\partial\mathcal{V}_{\Upsilon_{\delta}}$,
\item $H_{\delta}^{\circ}:\overline{\mathcal{V}}_{\Upsilon_{\delta}^{*}}\to\mathbb{R}$
is discrete harmonic, with $H_{\delta}^{\circ}:=F_{\delta}^{\circ}$
on $\partial\mathcal{V}_{\Upsilon_{\delta}^{*}}$,
\item $S_{\delta}^{\circ}:\overline{\mathcal{V}}_{\Upsilon_{\delta}^{*}}\to\mathbb{R}$
is discrete superharmonic, with $S_{\delta}^{\circ}:=0$ on $\partial\mathcal{V}_{\Upsilon_{\delta}^{*}}$.
\end{itemize}
Let us further decompose $H_{\delta}^{\bullet}$ as $A_{\delta}^{\bullet}+B_{\delta}^{\bullet}$,
where 
\begin{itemize}
\item $A_{\delta}^{\bullet}$ is discrete harmonic with $A_{\delta}^{\bullet}:=F_{\delta}\left(\partial\Omega\right)$
on $\partial\mathcal{V}_{\Omega_{\delta}}$ and $A_{\delta}^{\bullet}:=H_{\delta}^{\bullet}$
on $\partial\mathcal{V}_{\Upsilon_{\delta}}\setminus\partial\mathcal{V}_{\Omega_{\delta}}$.
\item $B_{\delta}^{\bullet}$ is discrete harmonic with $B_{\delta}^{\bullet}:=H_{\delta}^{\bullet}-F_{\delta}\left(\partial\Omega\right)$
on $\partial\mathcal{V}_{\Omega_{\delta}}$ and $B_{\delta}^{\bullet}:=0$
on $\partial\mathcal{V}_{\Upsilon_{\delta}}\setminus\partial\mathcal{V}_{\Omega_{\delta}}$.
\end{itemize}
The situation is hence the following: for any $z\in\mathcal{V}_{\Omega_{\delta}}$
and $w\in\mathcal{V}_{\Omega_{\delta}^{*}}$ such that $\left|z-w\right|=\delta/\sqrt{2}$,
from the construction of $F_{\delta}$, the superharmonicity of $F_{\delta}^{\circ}$
and the subharmonicity of $F_{\delta}^{\bullet}$, we have
\begin{equation}
H_{\delta}^{\circ}\left(w\right)\leq F_{\delta}^{\circ}\left(w\right)\leq F_{\delta}^{\bullet}\left(z\right)\leq H_{\delta}^{\bullet}\left(z\right)=A_{\delta}^{\bullet}\left(z\right)+B_{\delta}^{\bullet}\left(z\right).\label{eq:sandwich}
\end{equation}

It follows easily from Remark \ref{rem:integral-square} that as $n\to\infty$,
we have that $F_{\delta_{n}}\to F$, uniformly on the compact subsets
of $\Omega\setminus\left\{ a\right\} $ (since $f_{\delta_{n}}\to f$).

Let us now check that $F$ satisfies the conditions of Lemma \ref{lem:uniqueness-continuous-observable}.
$H_{\delta_{n}}^{\circ}$ and $H_{\delta_{n}}^{\bullet}$ are uniformly
close to each other on $\partial\Upsilon\setminus\partial\Omega$
(they are equal to $F_{\delta_{n}}^{\circ}$ and $F_{\delta_{n}}^{\bullet}$
there, and these functions are uniformly close to each other near
$\partial\Upsilon\setminus\partial\Omega$, as follows easily from
the convergence of $f_{\delta_{n}}$). To control $B_{\delta_{n}}$,
we use the following lemma, which is proven at the end of the section.
\begin{lem}
\label{lem:small-harmonic-function}As $n\to\infty$, $B_{\delta_{n}}^{\bullet}\to0$
uniformly on the compact subsets of $\Upsilon$. 
\end{lem}
Observe that $F_{\delta_{n}}\left(\partial\Omega\right)$ is uniformly
bounded. Suppose indeed that it would not be the case and (by extracting
a subsequence) that $F_{\delta_{n}}\left(\partial\Omega\right)\to\infty$
(say). We would have $H_{\delta_{n}}^{\circ}\to\infty$, since $H_{\delta_{n}}^{\circ}$
is harmonic and bounded on $\partial\Upsilon\setminus\partial\Omega$
(since it is equal to $F_{\delta_{n}}^{\circ}$ there). We also would
have $A_{\delta_{n}}^{\bullet}\to\infty$, for the same reasons. By
Equation \ref{eq:sandwich} and Lemma \ref{lem:small-harmonic-function},
it would imply that $F_{\delta_{n}}$ would blow up on $\Upsilon$,
which would contradict the fact that it converges uniformly to $F$
on the compact subsets of $\Upsilon$. 

We deduce that $H_{\delta_{n}}^{\circ}$ and $A_{\delta_{n}}^{\bullet}$
are uniformly bounded on $\overline{\Upsilon}$.

We have that $H_{\delta_{n}}^{\circ}\to F$ and $A_{\delta_{n}}^{\bullet}\to F$
as $n\to\infty$, uniformly on the compact subsets of $\Upsilon$.
From the discrete Beurling estimate (see \cite{kesten}) and the uniform
boundedness of $H_{\delta_{n}}^{\circ}$ and $A_{\delta_{n}}^{\bullet}$
near $\partial\Omega$, we readily obtain

\begin{eqnarray*}
\limsup_{n\to\infty}\left|A_{\delta_{n}}^{\bullet}\left(z\right)-F_{\delta_{n}}\left(\partial\Omega\right)\right| & \underset{z\to\partial\Omega}{\longrightarrow} & 0,\\
\limsup_{n\to\infty}\left|H_{\delta_{n}}^{\circ}\left(z\right)-F_{\delta_{n}}\left(\partial\Omega\right)\right| & \underset{z\to\partial\Omega}{\longrightarrow} & 0,
\end{eqnarray*}
and we deduce that $F$ continuously extends to $\partial\Omega$
and is constant there. 

To show that $\partial_{\nu_{\mathrm{out}}\left(z\right)}F\leq0$
for all $z\in\partial\Omega$, we consider the harmonic conjugate
$C$: it is the unique function $C$ (defined on the universal cover
of $\Omega\setminus\left\{ a\right\} $ and normalized to be $0$
at an arbitrary interior point $x$) such that $F+iC$ is holormophic.
By the Cauchy-Riemann equations, we have 
\[
\partial_{\nu_{\mathrm{out}}\left(z\right)}F=\partial_{\tau_{\mathrm{ccw}}\left(z\right)}C\,\,\,\,\forall z\in\partial\Omega,
\]
where $\partial_{\tau_{\mathrm{ccw}}\left(z\right)}$ is the tangential
derivative on $\partial\Omega$ in counterclockwise direction and
the condition $\partial_{\nu_{\mathrm{out}}\left(z\right)}F\leq0$
becomes $\partial_{\tau_{\mathrm{ccw}}\left(z\right)}C\leq0$. This
latter condition is equivalent to the one that $C$ is non-increasing
when going counterclockwise along (the universal cover of) $\partial\Omega$. 

Let us now check that this condition is satisfied. Take $\Upsilon$
as before and denote by $\tilde{\Upsilon}$ its universal cover.

For each $\delta>0$, let $C_{\delta}^{\circ}:\overline{\mathcal{V}}_{\tilde{\Upsilon}_{\delta}^{*}}\to\mathbb{R}$
be the discrete harmonic conjugate of $H_{\delta}^{\bullet}$ (lifted
to $\mathcal{V}_{\tilde{\Upsilon}_{\delta}}$), defined by integrating
the discrete Cauchy-Riemann equations 
\begin{eqnarray*}
\overline{\partial}_{\delta}\left(H_{\delta}^{\bullet}+iC_{\delta}^{\circ}\right)\left(z\right) & = & 0\,\,\,\,\forall z\in\mathcal{V}_{\tilde{\Upsilon}_{\delta}^{M}},
\end{eqnarray*}
and with the normalization $C_{\delta}^{\circ}\left(x\right)=0$.
By subharmonicity of $F_{\delta}^{\bullet}$, we have $F_{\delta}^{\bullet}\leq H_{\delta}^{\bullet}$
on $\mathcal{V}_{\Upsilon_{\delta}^{M}}$ and hence, since $F_{\delta}^{\bullet}=H_{\delta}^{\bullet}$
on $\partial\mathcal{V}_{\Omega_{\delta}^{M}}$, 
\[
\partial_{\nu_{\mathrm{out}}\left(z\right)}H_{\delta}^{\bullet}\leq\partial_{\nu_{\mathrm{out}}\left(z\right)}F_{\delta}^{\bullet}\leq0\,\,\,\,\forall z\in\partial\mathcal{V}_{\Omega_{\delta}^{M}},
\]
and we deduce by the discrete Cauchy-Riemann equations that $C_{\delta}^{\circ}$
is non-increasing when going along the universal cover of $\partial\mathcal{V}_{\Omega_{\delta}^{*}}$
in counterclockwise direction. 

On the compact subsets of $\tilde{\Upsilon}$, since the (normalized)
discrete derivatives of $H_{\delta_{n}}^{\bullet}$ converge uniformly
(see Remark 3.2 in \cite{chelkak-smirnov-i}) to the derivatives of
$F$, it is easy to check that $C_{\delta_{n}}^{\circ}$ also converges
uniformly to $C$. Since $C_{\delta}^{\circ}$ is non-increasing (when
going along the universal cover of $\partial\Omega$), we have that
$C_{\delta_{n}}^{\circ}$ is locally uniformly bounded (uniformly
with respect to $n$) on the universal cover of $\partial\Omega$
(if it would blow up there as $n\to\infty$, it would also blow up
on $\tilde{\Upsilon}$), and hence it is bounded everywhere on the
closure of $\tilde{\Upsilon}$. 

From there, we deduce that $C$ is non-increasing on the (counterclockwise-oriented)
universal cover of $\partial\Omega$: if it would not be the case,
using again the discrete Beurling estimate \cite{kesten}, we would
obtain a contradiction (in the $n\to\infty$ limit) with the fact
that $C_{\delta_{n}}^{\circ}$ is non-decreasing.
\end{proof}

\begin{proof}[Proof of Lemma \ref{lem:small-harmonic-function}]
 For $z\in\partial\mathcal{V}_{\Upsilon_{\delta}}$, let us write
$P_{\delta}\left(z,\cdot\right):\overline{\mathcal{V}}_{\Upsilon_{\delta}}\to\mathbb{R}$
for the discrete harmonic function such that $P_{\delta}\left(z,\cdot\right)=\mathbf{1}_{\left\{ z\right\} }\left(\cdot\right)$
on $\partial\mathcal{V}_{\Upsilon_{\delta}}$ (this is the discrete
harmonic measure of $\left\{ z\right\} $). By uniqueness of the solution
to the discrete Dirichlet problem, we can write
\[
B_{\delta}\left(y\right)=\sum_{z\in\partial\mathcal{V}_{\Omega_{\delta}}}B_{\delta}\left(z\right)P\left(z,y\right)\,\,\,\,\forall y\in\mathcal{V}_{\Upsilon_{\delta}}.
\]
As $\delta\to0$, we have that $P_{\delta}\left(x,\cdot\right)\to0$
on the compact subsets of $\Upsilon$, uniformly with respect to $x$
(this follows directly from Proposition 2.11 in \cite{chelkak-smirnov-i}).
By the construction of $F_{\delta}$ and the boundary conditions (Propositions
\ref{pro:boundary-condition} and \ref{pro:constant-value-integral-square}),
we have 
\[
B_{\delta}\left(z\right)=F_{\delta}\left(z\right)-F_{\delta}\left(\partial\Omega\right)=\sqrt{2}\cos\left(\frac{3\pi}{8}\right)\left|f_{\delta}\left(m\right)\right|^{2}\delta
\]
 for any $z\in\partial\mathcal{V}_{\Omega_{\delta}}$, where $m\in\partial_{0}\mathcal{V}_{\Omega_{\delta}^{M}}$
is the midpoint of the edge between $z$ and its neighbor in $\mathcal{V}_{\Omega_{\delta}}$.
Since 
\[
\sum_{m\in\partial_{0}\mathcal{V}_{\Omega_{\delta_{n}}^{M}}}\left|f_{\delta_{n}}\left(m\right)\right|^{2}\delta_{n}.
\]
is uniformly bounded by Lemma \ref{lem:boundedness-bdry-integral-square},
we readily deduce that $B_{\delta_{n}}\to0$ uniformly on the compact
subsets of $\Upsilon$.
\end{proof}

\section*{Appendix A}

\begin{figure}
\includegraphics[scale=0.4]{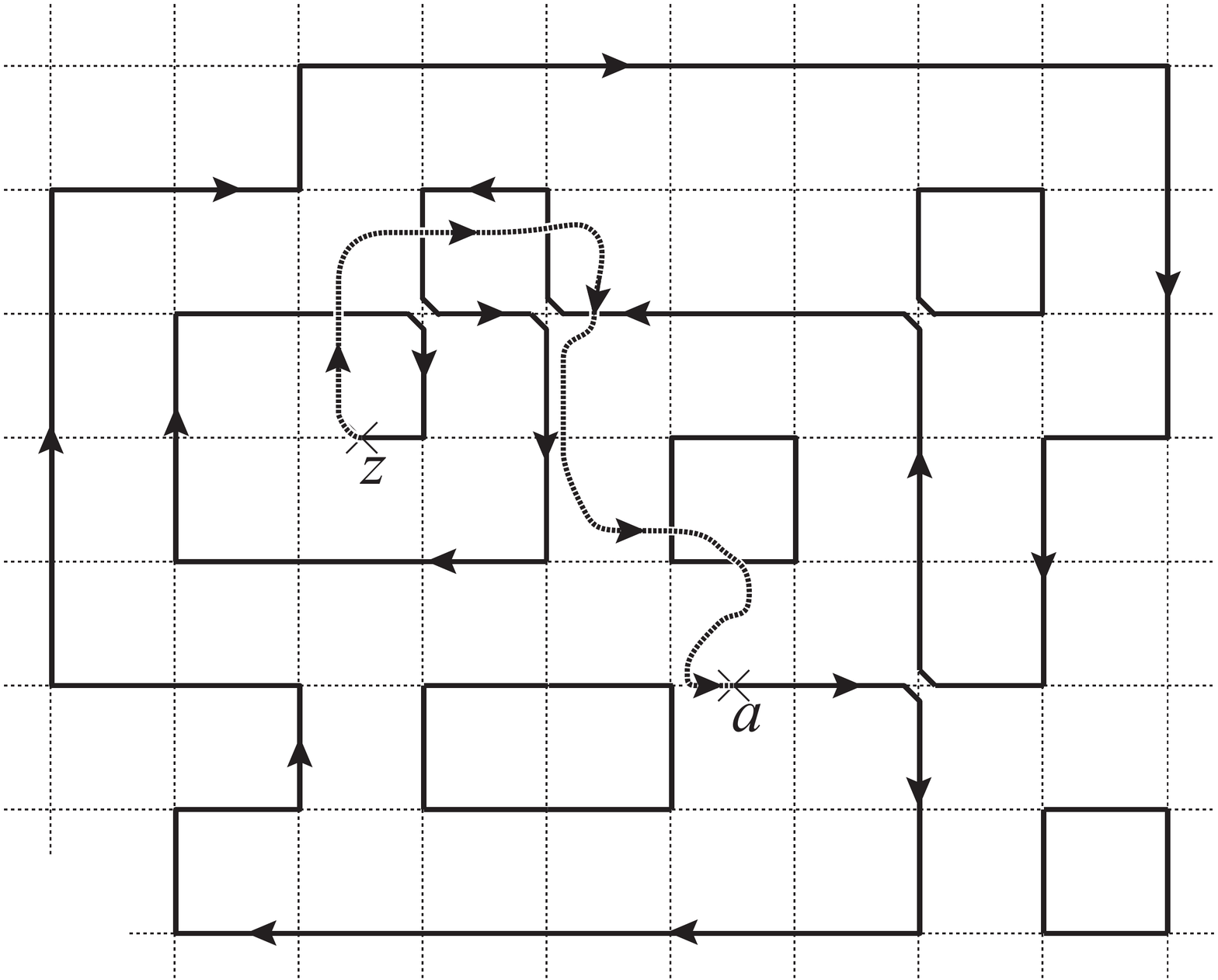}

\caption{\label{fig:configuration-loop}The oriented loop\emph{ $\mathcal{L}$}
formed by adding the curve $\mu$ (dotted) to $\lambda$. In this
case, $N_{1}=6$ and $N_{2}=4$. }
\end{figure}

\begin{figure}
\includegraphics[scale=0.4]{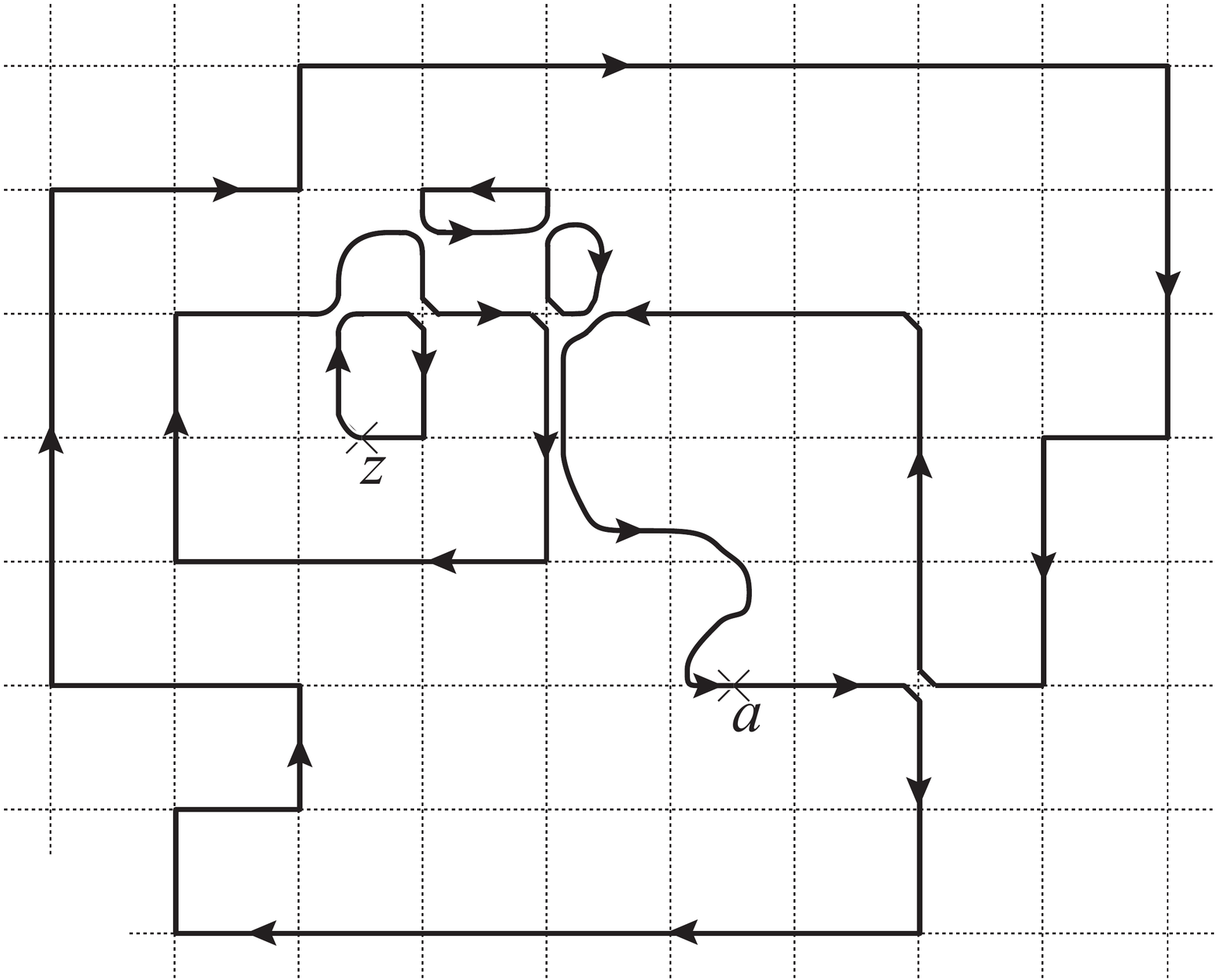}

\caption{\label{fig:configuration-loops}The five simple loops obtained from
$\mathcal{L}$ after four rearrangments (and discarding the loops
that were not part of $\lambda$).}
\end{figure}

We give here the proof of Lemma \ref{lem:well-definedness}: for a
configuration $\gamma\in\mathcal{C}_{\Omega_{\delta}}\left(a,z\right)$,
the winding (modulo $4\pi$) of an admissible walk on $\gamma$ (see
Figure \ref{fig:configuration-walk}) is independent of the choice
of that walk.
\begin{proof}[Proof of Lemma \ref{lem:well-definedness}]
Without loss of generality, half-edges of $\gamma$ emanate from
$z$ and $a$ in the same direction, so the winding is a multiple
of $2\pi$.

Add a curve $\mu$ from $z$ to $a$, which emanates in opposite direction
from $\gamma$ and run slightly off the lattice, so that $\mu$ is
transversal to $\gamma$ when an intersection occurs (see Figure \ref{fig:configuration-loop}).
Let $N_{1}$ be the number of intersections of $\mu$ with $\gamma$.

Take any admissible walk $\lambda$ along $\gamma$. The rest of $\gamma$
can be split into disjoint cycles. So, if $N_{2}$ is the number of
intersections of $\lambda$ with $\gamma$, then $N_{2}=N_{1}\,\,(\mathrm{mod}\,\,2)$.
Indeed, their difference comes from cycles, which are disjoint from
$\lambda$ and so intersect $\mu$ an even number of times (see Figure
\ref{fig:configuration-loop}).

The concatenation of $\lambda$ and $\mu$ (when oriented) forms a
loop $\mathcal{L}$, which has several intersections (when $\lambda$
and $\mu$ run transversally). At each of those, change the connection
so that there is no intersection, but instead two turns -- one left
and one right. Each of $N_{2}$ rearrangements either adds or removes
one loop, so after the procedure, $\mathcal{L}$ splits into $N_{3}$
simple loops with $N_{3}=N_{2}\,\,(\mathrm{mod}\,\,2)$ (see Figure
\ref{fig:configuration-loops}).

Each of the $N_{3}$ simple loops has winding $2\pi$ or $-2\pi$,
so $\mathbf{W}\left(\mathcal{L}\right)=2\pi N_{3}=2\pi N_{1}\,\,(\mathrm{mod}\,\,4\pi)$.

We conclude that, mod $4\pi$, $\mathbf{W}(\lambda)=\mathbf{W}(\mathcal{L})-\mathbf{W}(\mu)=N_{1}-\mathbf{W}(\mu)$
and so $\mathbf{W}(\lambda)\,\,\left(\mathrm{mod}\,\,4\pi\right)$
is independent of its particular choice.
\end{proof}

\section*{Appendix B}

We prove here some technical results concerning the discrete full-plane
spinor, introduced in Section \ref{sub:disc-hol-spin-full-plane}.
Let us denote by $C_{\delta}\left(\cdot,\cdot\right):=C_{0}\left(\frac{2\cdot}{\delta},\frac{2\cdot}{,\delta}\right)$
the rescaled version of Kenyon's coupling function (defined in \cite{kenyon-i}).
\begin{lem}
\label{lem:full-plane-obs-s-hol}With the notation and assumptions
of Proposition \ref{pro:full-plane-observable-s-holomorphicity},
the functions 
\begin{eqnarray*}
G_{1}:\mathcal{V}_{\mathbb{C}_{\delta}^{M}}\setminus\left\{ a\right\}  & \to & \mathbb{C}\\
z & \mapsto & e^{\frac{\pi i}{8}}\left(C_{\delta}\left(a+\frac{\delta}{2},z\right)+C_{\delta}\left(a-\frac{i\delta}{2},z\right)\right),\\
G_{2}:\mathcal{V}_{\mathbb{C}_{\delta}^{M}}\setminus\left\{ a\right\}  & \to & \mathbb{C}\\
z & \mapsto & e^{\frac{5\pi i}{8}}\left(C_{\delta}\left(a-\frac{\delta}{2},z\right)+C_{\delta}\left(a+\frac{i\delta}{2},z\right)\right)
\end{eqnarray*}
 are s-holomorphic.\end{lem}
\begin{proof}
Set $\eta:=e^{\frac{\pi i}{8}}$ and for any vertex $z$ and any $\mu\in\left\{ \pm1,\pm i\right\} $,
set $z_{\mu}:=z+\frac{\mu\delta}{2}$. By translation invariance,
we have
\begin{eqnarray*}
G_{1}\left(z\right) & = & \eta\left(C_{\delta}\left(a,z_{-1}\right)+C_{\delta}\left(a,z_{i}\right)\right),\\
G_{2}\left(z\right) & = & i\eta\left(C_{\delta}\left(a,z_{1}\right)+C_{\delta}\left(a,z_{-i}\right)\right),
\end{eqnarray*}
where, on the right hand sides, the two values of $C_{\delta}\left(a,\cdot\right)$
are orthogonal: one is purely real and the other purely imaginary.
Let $x,y\in\mathcal{V}_{\mathbb{C}_{\delta}^{M}\setminus\left\{ a\right\} }$
be two adjacent medial vertices, with $x$ being the midpoint of a
horizontal edge of $\mathcal{E}_{\Omega_{\delta}}$ and $y$ the midpoint
of a vertical one, and let $e:=\left\langle x,y\right\rangle \in\mathcal{E}_{\mathbb{C}_{\delta}^{M}}$.
Then, there are four possibilities for the line $\ell:=\ell\left(e\right)$: 
\begin{itemize}
\item If $\ell=\eta\mathbb{R}$, we have that $x=y+\frac{1+i}{2}\delta$
and 
\begin{eqnarray*}
\mathsf{P}_{\ell}\left[G_{1}\left(x\right)\right] & = & \eta C_{\delta}\left(a,x_{-1}\right)=\eta C_{\delta}\left(a,y_{i}\right)=\mathsf{P}_{\ell}\left[G_{1}\left(y\right)\right],\\
\mathsf{P}_{\ell}\left[G_{2}\left(x\right)\right] & = & i\eta C_{\delta}\left(a,x_{-i}\right)=i\eta C_{\delta}\left(a,y_{1}\right)=\mathsf{P}_{\ell}\left[G_{2}\left(y\right)\right].
\end{eqnarray*}
 
\item If $\ell=\overline{\eta}^{3}\mathbb{R}$, we have that $x=y-\frac{1+i}{2}\delta$
and 
\begin{eqnarray*}
\mathsf{P}_{\ell}\left[G_{1}\left(x\right)\right] & = & \eta C_{\delta}\left(a,x_{i}\right)=\eta C_{\delta}\left(a,y_{-1}\right)=\mathsf{P}_{\ell}\left[G_{1}\left(y\right)\right],\\
\mathsf{P}_{\ell}\left[G_{2}\left(x\right)\right] & = & i\eta C_{\delta}\left(a,x_{1}\right)=i\eta C_{\delta}\left(a,y_{-i}\right)=\mathsf{P}_{\ell}\left[G_{2}\left(y\right)\right].
\end{eqnarray*}

\item If $\ell=\overline{\eta}\mathbb{R}$, we have $x=y+\frac{1-i}{2}\delta$
and
\begin{eqnarray*}
\mathsf{P}_{\ell}\left[G_{1}\left(x\right)-G_{1}\left(y\right)\right] & = & \frac{\overline{\eta}}{\sqrt{2}}\left(C_{\delta}\left(a_{1},x_{-1}\right)+iC_{\delta}\left(a_{1},x_{i}\right)\right.\\
 &  & \left.-iC_{\delta}\left(a,y_{-1}\right)-C_{\delta}\left(a,y_{i}\right)\right)\\
 & = & \frac{i\overline{\eta}}{\sqrt{2}}\left(\overline{\partial}_{\delta}C_{\delta}\left(a,\cdot\right)\right)\left(y\right)=0,
\end{eqnarray*}
and similarly
\begin{eqnarray*}
\mathsf{P}_{\ell}\left[G_{2}\left(x\right)-G_{2}\left(y\right)\right] & = & \frac{\overline{\eta}}{\sqrt{2}}\left(-C_{\delta}\left(a,x_{1}\right)+iC_{\delta}\left(a,x_{-i}\right)\right.\\
 &  & \left.+C_{\delta}\left(a,y_{-i}\right)-iC_{\delta}\left(a,y_{1}\right)\right).
\end{eqnarray*}
 
\item If $\ell=\eta^{3}\mathbb{R}$, we have that $x=y+\frac{i-1}{2}\delta$
and
\begin{eqnarray*}
\mathsf{P}_{\ell}\left[G_{1}\left(x\right)-G_{1}\left(y\right)\right] & = & \frac{1}{\sqrt{2}}\left(C_{\delta}\left(a,x_{-1}\right)-iC_{\delta}\left(a,x_{i}\right)\right.\\
 &  & \left.+iC_{\delta}\left(a,y_{-1}\right)-C_{\delta}\left(a,y_{i}\right)\right)\\
 & = & -\frac{\eta^{3}}{\sqrt{2}}\left(\overline{\partial}_{\delta}C_{\delta}\left(a,\cdot\right)\right)\left(x\right)=0.
\end{eqnarray*}
and similarly
\begin{eqnarray*}
\mathsf{P}_{\ell}\left[G_{2}\left(x\right)-G_{2}\left(y\right)\right] & = & \frac{\eta^{3}}{\sqrt{2}}\left(-C_{\delta}\left(a,x_{1}\right)-iC_{\delta}\left(a,x_{-i}\right)\right.\\
 &  & \left.+C_{\delta}\left(a,y_{-i}\right)+iC_{\delta}\left(a,y_{1}\right)\right)\\
 & = & \frac{i\eta^{3}}{\sqrt{2}}\left(\overline{\partial}_{\delta}C_{\delta}\left(a,\cdot\right)\right)\left(y\right)=0.
\end{eqnarray*}

\end{itemize}
This concludes the proof of the lemma. 
\end{proof}
We now turn to the singularity of $f_{\mathbb{C}_{\delta}}$ (Proposition
\ref{pro:full-plane-observable-singularity}):
\begin{prop}
\label{prop:disc-singularity-full-plane-obs}Near the midpoint of
a horizontal edge $a\in\mathcal{V}_{\mathbb{C}_{\delta}}$, for $x\in\left\{ \pm1,\pm i\right\} $,
set $a_{x}:=a+\frac{x\delta}{2}\in\mathcal{V}_{\mathbb{C}_{\delta}^{M}}$
and by $e_{x}:=\left\langle a,a_{x}\right\rangle \in\mathcal{E}_{\mathbb{C}_{\delta}^{M}}$.
Then the function $f_{\mathbb{C}_{\delta}}\left(a,\cdot\right)$ satisfies
the relations 
\begin{eqnarray*}
\mathsf{P}_{\ell\left(e_{1+i}\right)}\left[f_{\mathbb{C}_{\delta}}\left(a,a\right)\right] & = & \mathsf{P}_{\ell\left(e_{1+i}\right)}\left[f_{\mathbb{C}_{\delta}}\left(a,a_{1+i}\right)\right],\\
\mathsf{P}_{\ell\left(e_{1-i}\right)}\left[f_{\mathbb{C}_{\delta}}\left(a,a\right)\right] & = & \mathsf{P}_{\ell\left(e_{1-i}\right)}\left[f_{\mathbb{C}_{\delta}}\left(a,a_{1-i}\right)\right],\\
\mathsf{P}_{\ell\left(e_{-1+i}\right)}\left[f_{\mathbb{C}_{\delta}}\left(a,a\right)-1\right] & = & \mathsf{P}_{\ell\left(e_{-1+i}\right)}\left[f_{\mathbb{C}_{\delta}}\left(a,a_{-1+i}\right)\right],\\
\mathsf{P}_{\ell\left(e_{-1-i}\right)}\left[f_{\mathbb{C}_{\delta}}\left(a,a\right)-1\right] & = & \mathsf{P}_{\ell\left(e_{-1-i}\right)}\left[f_{\mathbb{C}_{\delta}}\left(a,a_{-1-i}\right)\right].
\end{eqnarray*}
\end{prop}
\begin{proof}
Set $\mathrm{c}:=\cos\left(\frac{\pi}{8}\right)$ and $\mathrm{s}:=\sin\left(\frac{\pi}{8}\right)$
and $\eta:=e^{\frac{i\pi}{8}}$. Exact values of the coupling function
$C_{0}$ that can be found in \cite{kenyon-i} (see Figure 6 there)
\begin{eqnarray*}
C_{0}\left(0,1\right)=-C_{0}\left(0,-1\right) & = & \phantom{-}\frac{1}{4},\\
C_{0}\left(0,i\right)=-C_{0}\left(0,-i\right) & = & -\frac{i}{4},\\
C_{0}\left(0,2+i\right)=C_{0}\left(0,-2+i\right) & = & -i\left(\frac{1}{\pi}-\frac{1}{4}\right),\\
C_{0}\left(0,1+2i\right)=C_{0}\left(1-2i\right) & = & \frac{1}{\pi}-\frac{1}{4},\\
C_{0}\left(0,2-i\right)=C_{0}\left(0,-2-2i\right) & = & i\left(\frac{1}{\pi}-\frac{1}{4}\right),\\
C_{0}\left(0,-1-2i\right)=C_{0}\left(0,-1+2i\right) & = & \frac{1}{4}-\frac{1}{\pi}.
\end{eqnarray*}
Using these values and the definition of $f_{\mathbb{C}_{\delta}}$,
a straightforward computation gives
\begin{eqnarray*}
f_{\mathbb{C}_{\delta}}\left(a,a_{1+i}\right) & = & \eta\left(\mathrm{c}\left(\frac{2}{\pi}-\frac{1+i}{2}\right)-i\mathrm{s}\left(-\frac{2i}{\pi}+\frac{1+i}{2}\right)\right),\\
f_{\mathbb{C}_{\delta}}\left(a,a_{1-i}\right) & = & \eta\left(\mathrm{c}\left(\frac{1+i}{2}\right)-i\mathrm{s}\left(\frac{2i+2}{\pi}-\frac{1+i}{2}\right)\right),\\
f_{\mathbb{C}_{\delta}}\left(a,a_{-1+i}\right) & = & \eta\left(\mathrm{c}\left(-\frac{2+2i}{\pi}+\frac{1+i}{2}\right)+i\mathrm{s}\left(\frac{1+i}{2}\right)\right),\\
f_{\mathbb{C}_{\delta}}\left(a,a_{-1-i}\right) & = & \eta\left(\mathrm{c}\left(\frac{2i}{\pi}-\frac{1+i}{2}\right)+i\mathrm{s}\left(\frac{2}{\pi}-\frac{1+i}{2}\right)\right).
\end{eqnarray*}
If we compute the projections of these values on the lines associated
with the medial edges $e_{x}$, a straightforward computation gives
\begin{eqnarray*}
\mathsf{P}_{\overline{\eta}^{3}\mathbb{R}}\left[f_{\mathbb{C}_{\delta}}\left(a,a_{1+i}\right)\right] & = & \frac{\overline{\eta}^{3}\mathrm{c}}{\sqrt{2}}=\mathsf{P}_{\overline{\eta}^{3}\mathbb{R}}\left[\frac{2+\sqrt{2}}{4}\right],\\
\mathsf{P}_{\eta^{3}\mathbb{R}}\left[f_{\mathbb{C}_{\delta}}\left(a,a_{1-i}\right)\right] & = & \frac{\eta^{3}\mathrm{c}}{\sqrt{2}}=\mathsf{P}_{\eta^{3}\mathbb{R}}\left[\frac{2+\sqrt{2}}{4}\right],\\
\mathsf{P}_{\overline{\eta}\mathbb{R}}\left[f_{\mathbb{C}_{\delta}}\left(a,a_{-1+i}\right)\right] & = & -\frac{\overline{\eta}\mathrm{s}}{\sqrt{2}}=\mathsf{P}_{\overline{\eta}\mathbb{R}}\left[\frac{2+\sqrt{2}}{4}-1\right],\\
\mathsf{P}_{\eta\mathbb{R}}\left[f_{\mathbb{C}_{\delta}}\left(a,a_{-1-i}\right)\right] & = & -\frac{\eta\mathrm{s}}{\sqrt{2}}=\mathsf{P}_{\eta\mathbb{R}}\left[\frac{2+\sqrt{2}}{4}-1\right].
\end{eqnarray*}
which is the desired result.
\end{proof}
Let us now recall the result of Kenyon concerning the convergence
of the function $C_{0}:$
\begin{thm}[Theorem 11 in \cite{kenyon-i}]
\label{thm:kenyon-cv-theorem}As $\left|z\right|\to\infty$, we have
\[
C_{0}\left(0,z\right)=\begin{cases}
\Re\mathfrak{e}\left(\frac{1}{\pi z}\right)+O\left(\frac{1}{\left|z\right|^{2}}\right) & \,\,\,\, z=2m+\left(2n+1\right)i:m,n\in\mathbb{Z},\\
i\Im\mathfrak{m}\left(\frac{1}{\pi z}\right)+O\left(\frac{1}{\left|z\right|^{2}}\right) & \,\,\,\, z=\left(2m+1\right)+2ni:m,n\in\mathbb{Z}
\end{cases}
\]

\end{thm}
From there, we can prove Theorem \ref{thm:full-plane-spinor-convergence}:
\begin{proof}[Proof of Theorem \ref{thm:full-plane-spinor-convergence}]
By rescaling the lattice of the theorem above, one readily deduces
that 
\[
C_{0}\left(\frac{2}{\delta}\left(a+\frac{\delta}{2}\right),\frac{2}{\delta}z\right)+C_{0}\left(\frac{2}{\delta}\left(a-\frac{i\delta}{2}\right),\frac{2}{\delta}z\right)\longrightarrow\frac{1}{2\pi\left(z-a\right)}\,\,\delta\to0,
\]
uniformly on the sets $\left\{ \left(a,z\right):\left|a-z\right|\geq\epsilon\right\} $.
The proof of the theorem follows then from the definition of $f_{\mathbb{C}_{\delta}}$.\end{proof}


\begin{thebibliography}{References}
\bibitem[AsMc11]{assis-mccoy}M. Assis, B. M. McCoy, The energy density
of an Ising half-plane lattice, \emph{J. Phys. A:Math. Theor.} 44:095003,
2011.

\bibitem[Bax89]{baxter}R. Baxter, \emph{Exactly solved models in
statistical mechanics}. Academic Press Inc. {[}Harcourt Brace Jovanovich
Publishers{]}, London, 1989.

\bibitem[BoDT09]{boutillier-de-tiliere-i}C. Boutillier, B. de Tilière,
The critical Z-invariant Ising model via dimers: locality property.
\emph{Comm. Math. Phys}, to appear. arXiv:0902.1882v1, 2009.

\bibitem[BoDT08]{boutillier-de-tiliere-ii}C. Boutillier, B. de Tilière,
The critical Z-invariant Ising model via dimers: the periodic case.
\emph{PTRF} 147:379-413, 2010. arXiv:0812.3848v1.

\bibitem[BuGu93]{burkhard-guim}T. Burkhardt, I. Guim, Conformal theory
of the two-dimensional Ising model with homogeneous boundary conditions
and with disordered boundary fields,\emph{ Phys. Rev. B} (1), 47:14306-14311,
1993.

\bibitem[Car84]{cardy-i}J. Cardy, Conformal invariance and Surface
Critical Behavior\emph{, Nucl. Phys.} B 240:514-532, 1984.

\bibitem[ChSm08]{chelkak-smirnov-i}D. Chelkak, S. Smirnov, Discrete
complex analysis on isoradial graphs\emph{. Adv. in Math.}, to appear.
arXiv:0810.2188v1.

\bibitem[ChSm09]{chelkak-smirnov-ii}D. Chelkak, S. Smirnov, Universality
in the 2D Ising model and conformal invariance of fermionic observables\emph{.
Inv. Math., to appear}. arXiv:0910.2045v2, 2009.

\bibitem[CFL28]{courant-friedrichs-lewy}R. Courant, K. Friedrichs,
H. Lewy. Uber die partiellen Differenzengleichungen der mathematischen
Physik.\emph{ Math. Ann.}, 100:32-74, 1928.

\bibitem[DMS97]{di-francesco-mathieu-senechal}P. Di Francesco, P.
Mathieu, D. Sénéchal, \emph{Conformal Field Theory}, Graduate texts
in contemporary physics. Springer-Verlag New York, 1997.

\bibitem[Gri06]{grimmett}G. Grimmett, \emph{The Random-Cluster Model}.
Volume 333 of Grundlehren der Mathematischen Wissenschaften, Springer-Verlag,
Berlin, 2006.

\bibitem[Isi25]{ising}E. Ising,\emph{ }Beitrag zur Theorie des Ferromagnetismus\emph{.
Zeitschrift für Physik,} 31:253-258, 1925.

\bibitem[KaCe71]{kadanoff-ceva}L. Kadanoff, H. Ceva, Determination
of an operator algebra for the two-dimensional Ising model\emph{.
Phys. Rev. B} (3), 3:3918-3939, 1971.

\bibitem[Kau49]{kaufman}B. Kaufman, Crystal statistics. II. Partition
function evaluated by spinor analysis\emph{. Phys. Rev., II. Ser.,}
76:1232-1243, 1949.

\bibitem[KaOn49]{kaufman-onsager-i}B. Kaufman, L. Onsager,\emph{
}Crystal statistics. III. Short-range order in a binary Ising lattice\emph{.
Phys. Rev., II.} Ser., 76:1244-1252, 1949.

\bibitem[Ken00]{kenyon-i}R. Kenyon, Conformal invariance of domino
tiling\emph{. Ann. Probab.}, 28:759-795, 2000.

\bibitem[Kes87]{kesten}H. Kesten, Hitting probabilities of random
walks on $\mathbb{Z}^{d}$,\emph{ Stochastic Process. Appl.} 25:164-187,
1987.

\bibitem[KrWa41]{kramers-wannier}H. A. Kramers and G. H. Wannnier.
Statistics of the two-dimensional ferromagnet. I.\emph{ Phys. Rev.
}(2), 60:252-262, 1941.

\bibitem[Len20]{lenz}W. Lenz, Beitrag zum Verständnis der magnetischen
Eigenschaften in festen Körpern\emph{. Phys. Zeitschr.,} 21:613-615,
1920.

\bibitem[Mer01]{mercat}C. Mercat, Discrete Riemann surfaces and the
Ising model\emph{. Comm. Math. Phys., }218:177-216, 2001.

\bibitem[McWu73]{mccoy-wu}B. M. McCoy, and T. T. Wu, \emph{The two-dimensional
Ising model}. Harvard University Press, Cambridge, Massachusetts,
1973.

\bibitem[Ons44]{onsager}L. Onsager, Crystal statistics. I.\emph{
}A two-dimensional model with an order-disorder transition\emph{,
Phys. Rev. (2),} 65:117-149, 1944.

\bibitem[Pal07]{palmer}J. Palmer, \emph{Planar Ising correlations}.
Birkhäuser, 2007.

\bibitem[Smi06]{smirnov-i}S. Smirnov\emph{, }Towards conformal invariance
of 2D lattice models\emph{. Sanz-Solé, Marta (ed.) et al., Proceedings
of the international congress of mathematicians (ICM), Madrid, Spain,
August 22--30, 2006. Volume II: Invited lectures, 1421-1451. Zürich:
European Mathematical Society (EMS),} 2006.

\bibitem[Smi10]{smirnov-ii}S. Smirnov\emph{, }Conformal invariance
in random cluster models. I. Holomorphic fermions in the Ising model.\emph{
Ann. Math. 172(2)}, 2010.

\bibitem[Yan52]{yang}C. N. Yang, The spontaneous magnetization of
a two-dimensional Ising model\emph{. Phys. Rev. (2)}, 85:808-816,
1952.\end{thebibliography}
\end{document}